\makeatletter\@ifundefined{selectfont}
{%
\documentstyle[12pt,epsf]{article}}
{\documentstyle[12pt,oldlfont,epsf]{article}}
\makeatother
\begin{document}\thispagestyle{empty} \begin{flushright}
OUT--4102--52 \\
hep-ph/yymmnnn \end{flushright}\vfill
\begin{center}{\large\bf
Matching QCD and HQET heavy--light currents \\[3pt]
at two loops and beyond }
\vfill{\bf
D.~J.~Broadhurst$^1$} and {\bf
A.~G.~Grozin$^2$}
\vglue 2mm
Physics Department,
Open University, \\
Milton Keynes, MK7 6AA, UK \end{center}\vfill

\noindent{\bf Abstract}
Heavy--light QCD currents are matched with HQET currents
at two loops and leading order in $1/m$.
A single formula applies to all current matchings.
As a by--product, a master formula for the two--loop anomalous dimension of
the QCD current $\bar{q}\gamma^{[\mu_1}\ldots\gamma^{\mu_n]}q$ is obtained,
yielding a new result for the tensor current.
The dependence of matching coefficients on
$\gamma_5$ prescriptions is elucidated.
Ratios of QCD matrix elements are obtained,
independently of the three--loop anomalous dimension of HQET currents.
The two--loop coefficient in
$f_{{\rm B}^*}/f_{\rm B}=1-2\alpha_{\rm s}(m_b)/3\pi
-K_b\alpha_{\rm s}^2/\pi^2+{\rm O}(\alpha_{\rm s}^3,1/m_b)$ is
\[K_b=\frac{83}{12}+\frac{4}{81}\pi^2+\frac{2}{27}\pi^2\log2-\frac19\zeta(3)
-\frac{19}{54}N_l+\Delta_c=6.37+\Delta_c\]
with $N_l=4$ light flavours, and a correction, $\Delta_c=0.18\pm0.01$,
that takes account of the non--zero ratio $m_c/m_b=0.28\pm0.03$.
Fastest apparent convergence would entail $\alpha_{\rm s}(\mu)$ at
$\mu=370$~MeV.
``Naive non--abelianization'' of large--$N_l$ results,
via $N_l\to N_l-\frac{33}{2}$,
gives reasonable approximations to exact two--loop results.
All--order results for anomalous dimensions and matching coefficients
are obtained at large $\beta_0=11-\frac23N_l$.
Consistent cancellation between infrared-- and ultraviolet--renormalon
ambiguities is demonstrated.
\vfill\begin{flushleft}
OUT--4102--52\\
6 October, 1994
\end{flushleft}
\footnoterule\noindent
$^1$) D.Broadhurst@open.ac.uk\\
$^2$) A.Grozin@open.ac.uk;
on leave of absence from the Budker Institute of Nuclear Physics,
Novosibirsk 630090, Russia

\newcommand{\arccoth}{\mathop{\rm arccoth}\nolimits}
\newcommand{\Li}{\mathop{\rm Li}\nolimits}
\newcommand{\LM}{\Lambda_{\overline{\rm MS}}}
\renewcommand{\theequation}{\thesection.\arabic{equation}}
\renewcommand{\hat}[1]{#1\llap{/}}
\newpage \setcounter{page}{1}

\section{Introduction}
\label{intro}

QCD problems with a single heavy quark staying approximately at rest are
conveniently described by an effective field theory---HQET~\cite{EH,GG}
(see~\cite{Neu} for review and references). QCD operators are expanded, in
powers of $1/m$, in terms of HQET operators, $m$ being the on--shell
heavy--quark mass. In this paper we study, at the two--loop level, the
relation between currents in the full theory and the effective theory.

Specifically, we consider heavy--light bilinear currents,
$J_0=\bar{q}_0\Gamma Q_0$, where $\Gamma$ is a Dirac matrix and the
subscript 0 denotes an unrenormalized quantity. The $\overline{\rm
MS}$--renormalized QCD current, $J(\mu)=Z_J^{-1}(\mu)J_0$, is expanded in
HQET operators as
\begin{equation}
J(\mu) = C_\Gamma(\mu)\tilde{J}(\mu)
+ \frac1{m} \sum_i B_i(\mu)\tilde{O}_i(\mu)
+ {\rm O}\left(\frac1{m^2}\right)\,,
\label{match}
\end{equation}
where $\tilde{J}(\mu)=\tilde{Z}_J^{-1}(\mu)\tilde{J}_0$ is the
corresponding renormalized HQET current,
$\tilde{J}_0=\bar{q}_0\Gamma\tilde{Q}_0$ is the unrenormalized HQET
current, $\tilde{Q}_0$ is a two--component static--quark field, satisfying
$\gamma_0\tilde{Q}_0=\tilde{Q}_0$, and $\tilde{O}_i(\mu)$ are dimension--4
HQET operators, with appropriate quantum numbers. The meaning of the
operator equality~(\ref{match}) is that on--shell matrix elements of
$J(\mu)$, in situations amenable to HQET treatment, after expansion to a
given order in $1/m$, coincide with on--shell matrix elements of the
right--hand side.

A single dimension--3 term appears on the right--hand side of~(\ref{match})
if the Dirac matrix $\Gamma$ satisfies the conditions
\begin{equation}
\gamma_0\Gamma=\sigma\Gamma\gamma_0 \quad (\sigma=\pm1), \quad
\gamma_\mu\Gamma\gamma^\mu=2\sigma h\Gamma\,,
\label{cond}
\end{equation}
where $h$ is a function of the space--time dimension, $d=4-2\varepsilon$.
For an antisymmetrized product of $n$ $\gamma$--matrices,
$\Gamma=\gamma^{[\mu_1}\ldots\gamma^{\mu_n]}$, one obtains
$h=\eta(n-2+\varepsilon)$, with $\eta=-\sigma(-1)^n$.

It is natural to perform the matching at a scale $\mu\sim m$, where the
matching coefficient $C_\Gamma(\mu)$ contains no large logarithm. One can
then use the renormalization group to relate QCD and HQET currents
renormalized at arbitrary scales, $\mu$ and $\tilde{\mu}$:
\begin{equation}
\exp\left(-\int\limits_{\alpha_{\rm s}(m)}^{\alpha_{\rm s}(\mu)}
\frac{\gamma_J(\alpha)}{\beta(\alpha)}
\frac{{\rm d}\alpha}{\alpha}
\right) J(\mu) = C_\Gamma(m)
\exp\left(-\int\limits_{\alpha_{\rm s}(m)}^{\alpha_{\rm s}(\tilde{\mu})}
\frac{\tilde{\gamma}_J(\alpha)}{\beta(\alpha)}
\frac{{\rm d}\alpha}{\alpha}
\right) \tilde{J}(\tilde{\mu}) + {\rm O}\left(\frac1{m}\right)\,,
\label{RG}
\end{equation}
where $\gamma_J={\rm d}\log Z_J/{\rm d}\log\mu$ and $\tilde{\gamma}_J={\rm
d}\log \tilde{Z}_J/{\rm d}\log\mu$ are the QCD and HQET current anomalous
dimensions, and $\beta=-{\rm d}\log\alpha_{\rm s}/{\rm d}\log\mu$.

To calculate the matching coefficient $C_\Gamma(\mu)$, we consider an
on--shell QCD matrix element, $M(\mu)= \left(Z_Q^{\rm os}Z_q^{\rm
os}\right)^{1/2} Z_J^{-1}(\mu)\Gamma_0$, obtained from the bare
proper--vertex function, $\Gamma_0$, by renormalizing the current and
performing on--shell wave--function renormalization. To zeroth order in
$1/m$, it is equal to $C_\Gamma(\mu)\tilde{M}(\mu)$, where
$\tilde{M}(\mu)=\left(\tilde{Z}_Q^{\rm os}Z_q^{\rm os}\right)^{1/2}
\tilde{Z}_J^{-1}(\mu)\tilde{\Gamma}_0$. Hence we obtain
\begin{equation}
C_\Gamma(\mu) = \left(\frac{Z_Q^{\rm os}}{\tilde{Z}_Q^{\rm os}}\right)^{1/2}
\frac{\tilde{Z}_J(\mu)}{Z_J(\mu)} \frac{\Gamma_0}{\tilde{\Gamma}_0}\,,
\label{main}
\end{equation}
with a $\mu$ dependence coming only from $Z_J(\mu)$ and $\tilde{Z}_J(\mu)$,
in agreement with~(\ref{RG}).

The matching coefficient $C_\Gamma(\mu)$, for any $\Gamma$ of the
form~(\ref{cond}), was obtained at the one--loop level by Eichten and
Hill~\cite{EH}. The $1/m$ suppressed matching coefficients $B_i(\mu)$
in~(\ref{match}) were obtained for vector and axial currents, at one loop,
in~\cite{GH,Neu1}. Here we shall find the two--loop correction to the
leading matching coefficient $C_\Gamma(\mu)$. It is required for a more
accurate extraction of QCD matrix elements, such as $f_{\rm B}$ and
$f_{{\rm B}^*}$, from HQET results obtained, for example, from lattice
simulations or sum rules.

As can be seen from~(\ref{RG}), it is in general necessary to use
three--loop anomalous dimensions, $\gamma_J$ and $\tilde{\gamma}_J$, in
conjunction with a two--loop matching coefficient, $C_\Gamma(m)$. Whilst
some QCD anomalous dimensions are known at the three--loop level, the
universal HQET current anomalous dimension, $\tilde{\gamma}_J$, is
currently known only at the one--~\cite{SV,PW} and
two--loop~\cite{JM,BG,Gim} levels. There are, however, three cogent reasons
for calculating $C_\Gamma(m)$ to two loops. First, previous
experience~\cite{GBGS,BGS} strongly suggests that two--loop finite effects
dominate, numerically, over three--loop anomalous dimensions, near a
heavy--quark mass--shell. Secondly, the HQET anomalous dimension is
independent of the Dirac structure $\Gamma$ and hence is absent from
important ratios of physical quantities, such as $f_{{\rm B}^*}/f_{{\rm
B}}$. Finally, the three--loop term in $\tilde{\gamma}_J$ may be calculated
in future.

Motivated by these considerations, we here compute $C_\Gamma(m)$, to two
loops, for an arbitrary Dirac structure of the form~(\ref{cond}). Section~2
gives our method and general result, for any $\Gamma$. In Section~3 we
consider ratios of matching coefficients. After elucidating the dependence
on $\gamma_5$ prescriptions, we show that the two--loop
correction to $f_{{\rm B}^*}/f_{{\rm B}}$, at zeroth order in $1/m$, is
comparable in size to each of the effects previously computed, namely the
one--loop correction of~\cite{EH} and the ${\rm O}(1/m)$
correction of~\cite{Neu2,Ball}. In Section~4, we calculate
anomalous dimensions and matching coefficients to all orders in
$\alpha_{\rm s}$, in the limit of a large number of flavours, $N_f$.
We extend recent analyses~\cite{BB,Bigi,NS} by obtaining the
$\overline{\rm MS}$ matching coefficient for an arbitrary $\Gamma$
in this limit, confirming the cancellation of renormalon ambiguities in
physical matrix elements~\cite{NS} and validating an additional consistency
condition. Section~5 gives a summary of our main results.

\section{Two--loop matching calculation}
\label{2loop} \setcounter{equation}{0}

\begin{figure}[p]
\epsffile{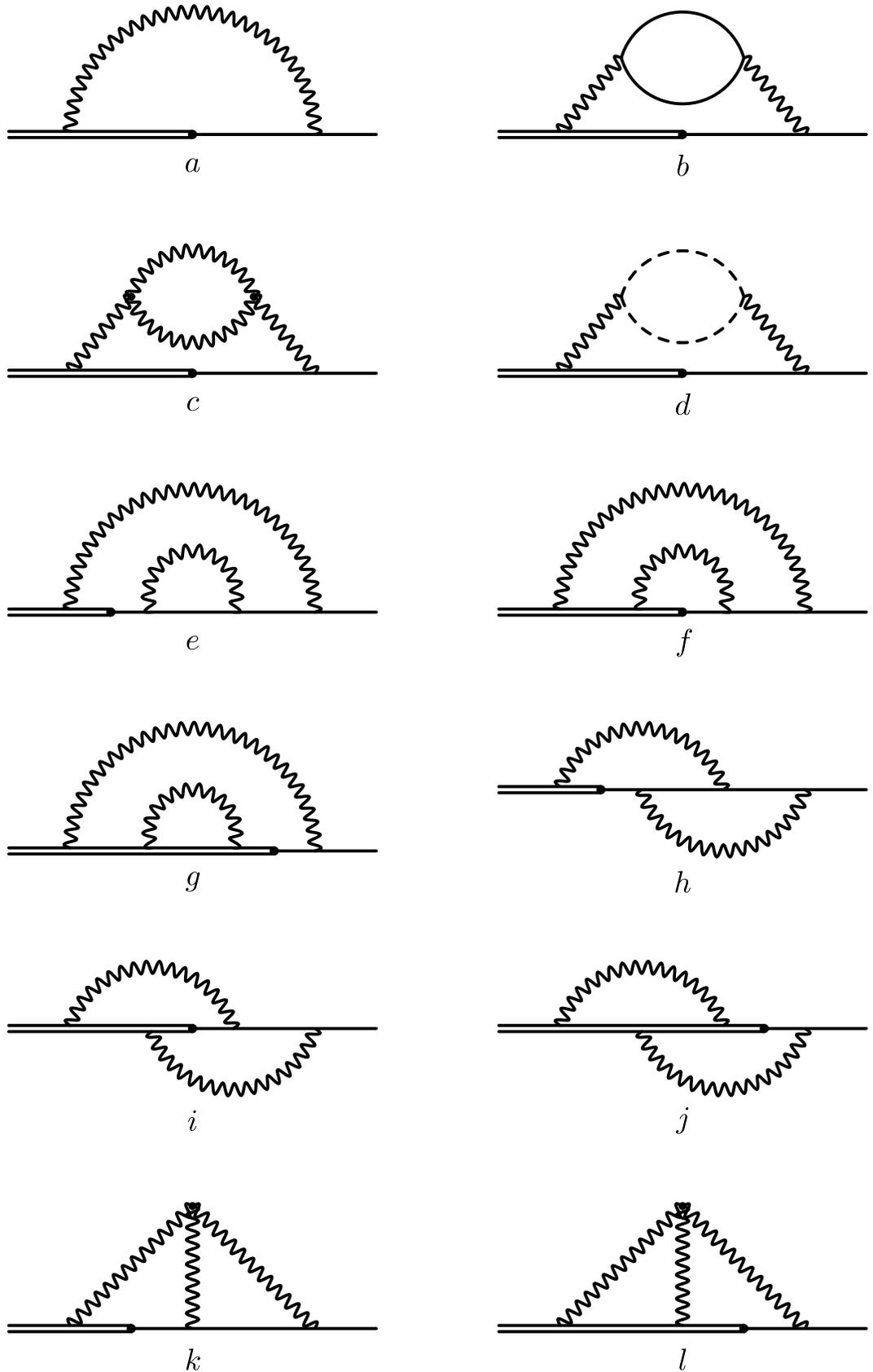}
\caption{One-- and two--loop diagrams for the proper vertex}
\label{Fig}
\end{figure}

The bare proper vertices, $\Gamma_0$ and $\tilde{\Gamma}_0$
in~(\ref{main}), may be evaluated at any on--shell momenta of the light and
heavy quarks. The calculation is greatly simplified by choosing the
momentum of each quark to vanish in HQET. This corresponds to a
heavy--quark momentum $p=m v$ in QCD, where $v=(1,\vec{0})$ in the
heavy--quark rest frame. One-- and two--loop diagrams for the vertex are
presented in Fig.~\ref{Fig}.

In the quark loop of diagram~\ref{Fig}b we include $N_f=N_l+1$ quark
flavours, one of which is the external heavy flavour, whilst the remaining
$N_l$ light flavours of quark have masses $m_i<m$, any of which may be
zero. In the case of an external $b$ quark, $N_f=4+1=5$ and only the
mass ratio $m_c/m_b\sim0.3$ is significantly greater than zero.

We stress that the loop containing the external heavy--quark flavour, with
mass $m_i=m$, must be included in all 6 terms on the right--hand side
of~(\ref{main}). This loop is certainly present in the QCD terms and we
must therefore include it the HQET terms, as well. The HQET modification of
QCD propagators occurs only along a single heavy--quark line, going through
all diagrams; HQET is blind to the contents of loops. In effect, HQET has
no memory that the static--quark line had one of the massive flavours that
still occur in loop corrections. Just as in conventional QCD, heavy loops
decouple from {\em physical\/} quantities, at low momenta; not from
renormalization constants, or bare vertices. We shall demonstrate the
physical decoupling of the $t$\/--quark loop from $f_{{\rm B}^*}/f_{\rm B}$
and show that the contribution of the $b$\/--quark loop is considerably
less than that of a light--quark loop.

We proceed to analyze, in turn, the 3 HQET quantities and 3 QCD quantities
on the right--hand side of~(\ref{main}).

\subsection{HQET calculation}

Here we deal with the HQET quantities $\tilde{\Gamma}_0$, $\tilde{Z}_J$,
and $\tilde{Z}_Q^{\rm os}$.

In the case of the HQET proper vertex, $\tilde{\Gamma}_0$, all the
dimensionally regularized diagrams of Fig.~\ref{Fig} vanish, for external
quarks with zero residual momenta. This is immediately obvious for all
except diagram~\ref{Fig}b, with a massive--quark loop, since no other
diagram contains a scale. In fact, this diagram vanishes too, because the
gluon propagator contains a quark--loop insertion and is hence transverse,
i.~e.\ proportional to $(g_{\mu\nu}-k_\mu k_\nu/k^2)$. This tensor is
contracted with $v^\nu$, from the heavy--quark vertex, and with
$\hat{k}\gamma^\mu/k^2$, from the light--quark vertex, $\gamma^\mu$,
adjacent to an internal light--quark propagator, $\hat{k}/k^2$, with the
same momentum as the gluon. The resulting contraction gives
$-\vec{\gamma}\cdot\vec{k}\gamma_0/k^2$, which is odd under reflection of
the spatial momentum $\vec{k}$ and hence vanishes when integrated over the
loop momentum $k$. Thus $\tilde{\Gamma}_0=\Gamma_{\rm B}$, where
$\Gamma_{\rm B}=\bar{u}_{q}\Gamma u_{Q}$ is merely the Born term.

The HQET current renormalization $\tilde{Z}_J$ does not depend on $\Gamma$.
In the $\overline{\rm MS}$ scheme, at two loops, it is fully determined by
the anomalous dimension~\cite{JM,BG,Gim}:
\begin{equation}
\tilde{\gamma}_J = -3 C_F \frac{\alpha_{\rm s}}{4\pi}
+ C_F \left(\frac{\alpha_{\rm s}}{4\pi}\right)^2
\left[C_F\left(\frac52-16\zeta(2)\right)
+C_A\left(-\frac{49}{6}+4\zeta(2)\right)
+\frac{10}{3}T_F N_f\right]\,,
\label{gjt}
\end{equation}
where $C_A=3$, $C_F=4/3$, $T_F=1/2$, in the case of QCD, and we omit
$\alpha_{\rm s}^3$ terms, here and subsequently, without further comment.

The HQET on--shell wave--function renormalization constant is defined by
$\tilde{Z}_Q^{\rm os}=\left(1 -({\rm d}\tilde{\Sigma}_0/{\rm
d}\omega)_{\omega=0}\right)^{-1}$ where $\tilde{\Sigma}_0(\omega)$ is the
bare static--quark self--energy, at residual energy $\omega$. As in the
case of the vertex, only diagrams containing a loop with a quark of
non--zero mass can contribute, since no other diagram contains a scale.
Thus we obtain
\begin{equation}
\left.\frac{{\rm d}\tilde{\Sigma}_0}{{\rm d}\omega}\right|_{\omega=0} =
-i C_F g_0^2 \int\frac{{\rm d}^d k}{(2\pi)^d}
\frac{v^\mu v^\nu}{(v k)^2}\frac{\Pi_{\mu\nu}(k)}{k^4} =
-i \frac{C_F g_0^2}{d-1} \int\frac{{\rm d}^d k}{(2\pi)^d}
\left(\frac{k^2}{(v k)^2}-1\right)
\frac{\Pi_\alpha^\alpha(k)}{k^6}\,,
\label{Sig0}
\end{equation}
where $\Pi_{\mu\nu}(k)= (g_{\mu\nu}-k_\mu
k_\nu/k^2)\Pi_\alpha^\alpha(k)/(d-1)$ gives the contribution of quark loops
to the polarization operator, whose trace depends only on $k^2$. We now
average over all directions of $k$, in $d$--dimensional space, obtaining
$<k^2/(v k)^2>\,=2-d$. A dimensionally regularized result is then readily
obtained from two--loop massive bubble integrals~\cite{GBGS}, which give
\begin{equation}
\left.\frac{{\rm d}\tilde{\Sigma}_0}{{\rm d}\omega}\right|_{\omega=0} =
-8 C_F T_F \frac{g_0^4}{(4\pi)^d}
\sum_{m_i>0} m_i^{-4\varepsilon}
\frac{(d-1)(d-6)\Gamma^2(1+\varepsilon)}{(d-2)(d-4)^2(d-5)(d-7)}\,,
\label{Sig}
\end{equation}
whose Laurent expansion yields the on--shell wave--function renormalization
constant
\begin{equation}
\tilde{Z}_Q^{\rm os}=1 + 2 C_F T_F
\left(\frac{\alpha_{\rm s}}{4\pi}\right)^2\!\sum_{m_i>0}
\left( \frac1{\varepsilon^2} - \frac4{\varepsilon}\log\frac{m_i}{\mu}
- \frac4{3\varepsilon} + 8\log^2\frac{m_i}{\mu}
+ \frac{16}3 \log\frac{m_i}{\mu} + \zeta(2) + \frac{26}9 \right).
\label{Zos}
\end{equation}
We note that~(\ref{Zos}) exhibits non--uniformity as $m_i\to0$: quarks with
masses $m_i>0$ are to be included; any quark with $m_i=0$ is excluded, by
dimensional regularization. This non--uniformity has the same infrared
origin as that observed in on--shell QCD wave--function
renormalization~\cite{BGS}, to which we now turn.

\subsection{QCD calculation}

For continuity, we deal with the QCD quantities, $\Gamma_0$, $Z_J$, and
$Z_Q^{\rm os}$, in reverse order.

A gauge--invariant on--shell wave--function renormalization constant,
$Z_Q^{\rm os}$, was obtained in $d$ dimensions, at the two--loop level,
in~\cite{BGS}. A quark flavour with a small but non--zero mass gives a
contribution to $Z_Q^{\rm os}$ different from that of a zero--mass flavour.
The same is seen to be true for $\tilde{Z}_Q^{\rm os}$ in~(\ref{Zos}). Each
discontinuity originates from the infrared region of small gluon momenta,
where HQET does not differ from QCD. Therefore these discontinuities should
cancel in the ratio $Z_Q^{\rm os}/\tilde{Z}_Q^{\rm os}$, which we indeed
find to have a smooth limit, as $m_i\to0$. This uniformity provides a
strong check of the massless and massive quark--loop contributions to
$Z_Q^{\rm os}$, obtained in~\cite{BGS}. The ratio
\begin{equation}
\frac{Z_Q^{\rm os}/\tilde{Z}_Q^{\rm os}}
{(Z_Q^{\rm os}/\tilde{Z}_Q^{\rm os})_{m_i=0}}
= 1 + 2 C_F T_F \left(\frac{\alpha_{\rm s}}{4\pi}\right)^2
\sum_{i=1}^{N_f} \Delta_Z\left(\frac{m_i}{m}\right)
\label{RZ}
\end{equation}
may be expressed as an integral of the difference,
$\Pi\left(-m_i^2/k^2\right)$, between the polarization operator with a
quark of mass $m_i$ in the loop and that with a massless quark. This
difference contains neither ultraviolet nor infrared divergences. The
integral over the gluon momentum $k$ is likewise well--behaved, and may be
performed in 4 dimensions. As is the case for all two--scale calculations
in this paper, the result may be expressed as a combination of three basic
dilogarithmic integrals, defined and evaluated in Appendix~A. Combining
$Z_Q^{\rm os}$, from~\cite{BGS}, with $\tilde{Z}_Q^{\rm os}$,
from~(\ref{Zos}), we obtain
\begin{equation}
\Delta_Z(r) = - 4\Delta_1(r) - 2\Delta_2(r) - 12\Delta_3(r)\,.
\label{DeltaZ}
\end{equation}

The QCD current renormalization constants $Z_J$ are known to two loops for
some, but not all, of the currents that we study. We shall simply apply a
minimal Ansatz for $Z_J$ and extract its coefficients from the requirement
of finiteness of~(\ref{main}), checking our resultant master formula for
$\gamma_J$ against known special cases.

The most difficult part of the problem is the calculation of the bare QCD
proper vertex, $\Gamma_0$. We shall first find it in the case when all
light flavours are massless, and include later the effects of finite
light--quark masses. Since we calculate the bare vertex on the
renormalized mass--shell,
$p=m v$, it is convenient to express the result in terms of the on--shell
mass $m$, treating the one--loop $d$\/--dimensional counterterm, $\Delta
m=m-m_0$, as a perturbation.

Initially, we make no assumption about the properties of the matrix
$\Gamma$, and note that each two--loop diagram for $\Gamma_0$ may be
written as a sum of terms, each of the form
\begin{equation}
\bar{u}_{q}\gamma_{\mu_1}\ldots\gamma_{\mu_l}\Gamma
\gamma_{\nu_1}\ldots\gamma_{\nu_r}u_{Q}\cdot
I^{\mu_1\ldots\mu_l;\nu_1\ldots\nu_r}\,,
\label{form0}
\end{equation}
where $I$ is some integral over loop momenta, $l$ is even, and $l+r\le8$.
After the integration, $I^{\mu_1\ldots\mu_l;\nu_1\ldots\nu_r}$ can contain
only $g^{\mu\nu}$ and $v^\alpha$. Resulting contractions of pairs of
$\gamma$--matrices on the left, and of pairs on the right, merely produce
additional terms of the form~(\ref{form0}), with smaller values of $l+r$.
Before performing the remaining contractions, one may anticommute
$\gamma$--matrices, so as to arrange that $\hat{v}$ occurs only on the
extreme left, or the extreme right, with the contracted indices in between
occurring in opposite orders on the left and right of $\Gamma$. Additional
terms, arising from anticommutators, have fewer $\gamma$--matrices, with
$l$ remaining even. Repeating this procedure for all values of $l+r$, from
8 down to 0, we may cast any diagram in the form
\begin{eqnarray}
&&D = \bar{u}_q \big[ \Gamma(x_0+x_1\hat{v})
+ \hat{v}\gamma_\alpha\Gamma\gamma^\alpha(x_2+x_3\hat{v})
+ \gamma_\alpha\gamma_\beta\Gamma\gamma^\beta\gamma^\alpha(x_4+x_5\hat{v})
\nonumber\\&&\quad{}
+ \hat{v}\gamma_\alpha\gamma_\beta\gamma_\gamma\Gamma
\gamma^\gamma\gamma^\beta\gamma^\alpha(x_6+x_7\hat{v})
+ \gamma_\alpha\gamma_\beta\gamma_\gamma\gamma_\delta\Gamma
\gamma^\delta\gamma^\gamma\gamma^\beta\gamma^\alpha x_8
\big] u_Q\,.
\label{form}
\end{eqnarray}

Now we assume that the matrix $\Gamma$ has the properties~(\ref{cond}). The
effect of each contraction is then to produce a factor $2\sigma h$. Terms
with an odd number of contractions necessarily contain $\hat{v}$ on the
left, which yields an extra $\sigma$ when moved to the right, where it
merely gives $\hat{v}u_Q=u_Q$. Thus we obtain a result involving only
powers of $h$:
\begin{equation}
D = \left[ (x_0+x_1) + (x_2+x_3)(2h) + (x_4+x_5)(2h)^2
+ (x_6+x_7)(2h)^3 + x_8(2h)^4 \right] \Gamma_{\rm B}\,.
\label{form1}
\end{equation}
We can find the coefficients $x_0+x_1$, $x_2+x_3$, $x_4+x_5$, $x_6+x_7$,
$x_8$, for each diagram, by taking, separately, a trace of the
$\gamma$--matrices on the light--quark line with $L_i$, and a trace on the
heavy--quark line with $H_i$, using the following 5 forms of $L_i\times
H_i$:
$1\times(1+\hat{v})$;
$\gamma_\mu\hat{v}\times(1+\hat{v})\gamma^\mu$;
$\gamma_\mu\gamma_\nu\times(1+\hat{v})\gamma^\nu\gamma^\mu$;
$\gamma_\mu\gamma_\nu\gamma_\rho\hat{v}\times
(1+\hat{v})\gamma^\rho\gamma^\nu\gamma^\mu$;
$\gamma_\mu\gamma_\nu\gamma_\rho\gamma_\sigma\times
(1+\hat{v})\gamma^\sigma\gamma^\rho\gamma^\nu\gamma^\mu$.
Performing the same operation on the generic form~(\ref{form}), we obtain 5
equations relating these double--traces to the desired coefficients.
Inverting these equations, once and for all, we may then convert the {\em
integrand\/} of any diagram into a 4th order polynomial in $h$, with
coefficients that are scalar functions of $v$ and the two loop momenta. For
reliability, we checked our general result, for each two--loop diagram,
against brute--force evaluation of 8 specific cases of Dirac matrix. The
general one--loop diagram~\ref{Fig}a, and its associated mass--counterterm
contribution, were similarly reduced to quadratic functions of $h$.

All the resulting one--loop scalar integrals can be expressed in terms of
\begin{equation}
I_0 = - \frac{i}{\pi^{d/2}} \int \frac{{\rm d}^d k}{k^2+2v k}
= \frac{\Gamma(1+\varepsilon)}{\varepsilon(1-\varepsilon)}\,.
\label{I0}
\end{equation}
The two--loop integrals may be reduced to combinations of $I_0^2$ and two
further terms:
\begin{eqnarray}
&&I_1 = - \frac1{\pi^d} \int
\frac{{\rm d}^d k {\rm d}^d l}{k^2(l-k)^2(l^2+2v l)}
\nonumber\\&&\quad
= \frac{1-4\varepsilon}{2\varepsilon^2(1-2\varepsilon)(1-\frac32\varepsilon)
(1-3\varepsilon)}
\frac{\Gamma(1+\varepsilon)\Gamma^2(1-\varepsilon)\Gamma(1+2\varepsilon)
\Gamma(1-4\varepsilon)}{\Gamma(1-2\varepsilon)\Gamma(1-3\varepsilon)}\,,
\nonumber\\
&&I_2 = - \frac1{\pi^d} \int
\frac{{\rm d}^d k {\rm d}^d l}{(k^2+2v k)(l^2+2v l)((k+l)^2+2v(k+l))}
\label{I12}\\&&\quad
= \frac{3(5d-18)(d-2)^2}{2(3d-8)(3d-10)(d-3)}I_0^2
- \frac{2(d-4)}{2d-7}I_1
- \frac{16(d-4)^2}{(3d-8)(3d-10)}I(\varepsilon)\,,
\nonumber
\end{eqnarray}
where~\cite{Br0,Br}
\begin{equation}
I(\varepsilon) = I + {\rm O}(\varepsilon), \quad
I = \pi^2\log2 - {\textstyle\frac32}\zeta(3)\,.
\label{I}
\end{equation}
To achieve this reduction, we used the package RECURSOR~\cite{Br}, written
in REDUCE~\cite{H} to implement recurrence relations derived from
integration by parts~\cite{GBGS,BGS}.

So as to have a strong check, we performed the calculation in an arbitrary
covariant gauge, verifying that the sum of all diagrams for the on--shell
vertex, including the mass counterterm, is gauge invariant in $d$
dimensions. The result can be written as
\begin{equation}
\frac{\Gamma_0}{\Gamma_{\rm B}} = 1
- C_F \frac{g_0^2 m^{-2\varepsilon}}{(4\pi)^{d/2}} I_0
\frac{(1-h)(d-2+2h)}{2(d-3)}
+ C_F \frac{g_0^4 m^{-4\varepsilon}}{(4\pi)^d}
\sum_{i=0}^3 \sum_{j=0}^2 \sum_{k=0}^3 a_{j i k} C_i J_j h_k\,,
\label{vert}
\end{equation}
in terms of 4 colour factors, 3 integral structures, and 4
current--specific functions. The colour factors are chosen as follows:
$C_0=C_F-\frac12C_A$; $C_1=C_F$; $C_2=T_F N_l$, from $N_l$ massless loops
in diagram~\ref{Fig}b; $C_3=T_F$, from the heavy--quark loop. The integrals
\begin{eqnarray}
&&J_0 = \frac{(d-2)I_0^2}{8(d-1)(d-3)^2(d-4)^2(d-5)(d-6)}\,,
\label{JI}\\
&&J_1 = \frac{I_1}{8(d-1)(d-3)(d-4)(2d-7)}\,,
\quad
J_2 = \frac{I_2}{4(d-1)(d-4)^2(d-6)}\,,
\nonumber
\end{eqnarray}
are chosen to make the coefficients $a_{j i k}$ polynomials in $d$. All
dependence on $\Gamma$ resides in $h_k=h^k$, for $k=0,1,2$, and
$h_3=h^3(h+d-4)$. Only diagrams~\ref{Fig}f and~\ref{Fig}i have sufficient
$\gamma$--matrices on each side of the vertex to generate $h^3$ and $h^4$
terms, and we find that each diagram produces the combination $h_3$, for
each integral structure. Moreover, the integral $J_2$ arises only from
diagram~\ref{Fig}j, with colour factor $C_0$, or from the heavy--quark loop
in diagram~\ref{Fig}b, with colour factor $C_3$. Thus several of the 48
coefficients $a_{j i k}$ vanish. The 28 that survive are listed in
Appendix~B.

We complete the calculation by including the effect of non--zero
light--quark masses. The difference of diagram~\ref{Fig}b, with a quark of
mass $m_i$ in the loop, and the same diagram with a massless quark,
contains neither ultraviolet nor infrared divergences. Therefore we
calculate it in 4 dimensions, obtaining a further term, to be added
to~(\ref{vert}), of the form
\begin{equation}
\frac{\Delta\Gamma_0}{\Gamma_{\rm B}}
= C_F T_F \left(\frac{\alpha_{\rm s}}{4\pi}\right)^2
\sum_{i=1}^{N_l} \Delta_\Gamma\left(\frac{m_i}{m}\right)\,,
\label{mass}
\end{equation}
where the sum runs over light flavours only, and $\Delta_\Gamma$ can be
reduced to the integrals~(\ref{Delta123}), as follows:
\begin{equation}
\Delta_\Gamma(r) = \frac43 \left[ - 2(1-h^2)\Delta_1(r)
- (1+2h-2h^2)\Delta_2(r) + 2h(2-h)\Delta_3(r) \right]\,,
\label{DeltaG}
\end{equation}
with $\Delta_\Gamma(0)=0$, by virtue of its definition. As a strong check
on the coefficients of the colour factors $C_3$ and $C_2$ in~(\ref{vert}),
coming from loops with quarks of masses $m_i=m$ and $m_i=0$, respectively,
we have verified that setting $d=4$ in their difference agrees with setting
$r=1$ in~(\ref{DeltaG}).

\subsection{Results}

We re--express the bare QCD vertex~(\ref{vert}) in terms of the
$\overline{\rm MS}$ coupling, $\alpha_{\rm s}(\mu)$, using one--loop
coupling--constant renormalization, with $N_f=N_l+1$ flavours, and expand
the result as a Laurent series in $\varepsilon$. Using $Z^{\rm os}_Q$
from~\cite{BGS}, $\tilde{\Gamma}_0=\Gamma_{\rm B}$, $\tilde{Z}_J(\mu)$
from~(\ref{gjt}), and $\tilde{Z}^{\rm os}_Q$ from~(\ref{Zos}), we obtain
the generic QCD current renormalization constant, $Z_J(\mu)$, by requiring
the finiteness of~(\ref{main}). The corresponding anomalous dimension is
that of the QCD current
$J(\mu)=\bar{q}\gamma^{[\mu_1}\ldots\gamma^{\mu_n]}q$, which does not
depend upon whether the quarks are heavy or light. We find that
\begin{eqnarray}
&&\gamma_J = - C_F \frac{\alpha_{\rm s}}{2\pi} (n-1)(n-3)
\left[1 + \frac{\alpha_{\rm s}}{4\pi}
\left(C_F\frac{5(n-2)^2-19}{2}-C_A\frac{3(n-2)^2-19}{3}\right)\right]
\nonumber\\
&&\quad{} - C_F \left(\frac{\alpha_{\rm s}}{4\pi}\right)^2 (n-1)(n-15)
\frac{11C_A-4T_F N_f}{9}\,,
\label{adim}
\end{eqnarray}
with a check being provided by the absence of the signature $\eta$, which
merely distinguishes different components of $J$. We reproduce the known
result for the scalar current~\cite{Tar}, with $n=0$, and, of course,
obtain a vanishing result for the vector current, with $n=1$. The two--loop
anomalous dimension of the $n=2$ current, $\bar{q}\sigma_{\mu\nu}q$,
appears to be absent from the literature, though it could be derived by
straightforward calculation in massless QCD. Instead, we have obtained
\begin{equation}
\left.\gamma_J\right|_{n=2} = C_F \frac{\alpha_{\rm s}}{2\pi}
\left[1 + \frac{\alpha_{\rm s}}{4\pi}
\left(-\frac{19}{2}C_F+\frac{257}{18}C_A
-\frac{26}{9}T_F N_f\right)\right]
\label{adim2}
\end{equation}
as a by--product of a more difficult massive on--shell calculation. The
currents with $n=3$ and $n=4$ are non--singlet axial and pseudoscalar
currents, with the 't~Hooft--Veltman $\gamma_5$, whose anomalous dimensions
are known to differ, at two loops~\cite{Tru,Lar} and beyond~\cite{Lar,LV},
from those with the naively anticommuting $\gamma_5$, with $n=1$ and $n=0$,
respectively. This difference is apparent in the final term
of~(\ref{adim}), which lacks invariance under $n\to4-n$. Our results for
$n=3$ and $n=4$ agree with~\cite{Tru} and~\cite{Lar}, respectively.

Finally we arrive at the general expression for the two--loop matching
coefficient
\begin{eqnarray}
&&C_\Gamma(m) = 1 + C_F \frac{\alpha_{\rm s}(m)}{4\pi}
\left[3(n-2)^2+(2-\eta)(n-2)-4\right]
\nonumber\\
&&\quad{} + C_F \left(\frac{\alpha_{\rm s}}{4\pi}\right)^2
\left[ C_F a_F + C_A a_A
+ T_F \sum_{i=1}^{N_f}
\left\{a_f+\Delta_J\left(\frac{m_i}{m}\right)\right\}\right]\,,
\label{cm}
\end{eqnarray}
where the sum now {\em includes\/} the heavy flavour, with $m_i=m$. The
one--loop term coincides with~\cite{EH}. The coefficients in the two--loop
term are
\begin{eqnarray}
&&a_F = \left(\frac{317}{24}-\frac{10}{3}\zeta(2)\right)(n-2)^4
+ 11(n-2)^3 - \frac{11}{2}\eta(n-2)^3
\nonumber\\&&\quad{}
+ \left(-\frac{253}{6}+48\zeta(2)-\frac{16}{3}I\right)(n-2)^2
- 2\eta(n-2)^2 - 20(n-2)
\nonumber\\&&\quad{}
+ \left(\frac{32}{3}-\frac{64}{3}\zeta(2)+\frac{8}{3}I\right)\eta(n-2)
+\frac{689}{16}-81\zeta(2)-8\zeta(3)+12I\,,
\nonumber\\
&&a_A = \left(-\frac{43}{12}+\frac{4}{3}\zeta(2)\right)(n-2)^4
- 2(n-2)^3 + \eta(n-2)^3
\label{coef}\\&&\quad{}
+ \left(\frac{9491}{216}-\frac{52}{3}\zeta(2)+\frac{8}{3}I\right)(n-2)^2
+ \frac{143}{18}(n-2)
\nonumber\\&&\quad{}
+ \left(-\frac{281}{18}+8\zeta(2)-\frac{4}{3}I\right)\eta(n-2)
-\frac{29017}{432}+29\zeta(2)+2\zeta(3)-6I\,,
\nonumber\\
&&a_f = \left(-\frac{445}{54}-\frac{8}{3}\zeta(2)\right)(n-2)^2
- \frac{2}{9}(n-2) + \frac{38}{9}\eta(n-2)
+ \frac{1745}{108}+\frac{20}{3}\zeta(2)\,.
\nonumber
\end{eqnarray}
The mass correction $\Delta_J(r)=\Delta_\Gamma(r)+\Delta_Z(r)$ depends only
on $h|_{\varepsilon=0}=\eta(n-2)$, because~(\ref{DeltaZ})
and~(\ref{DeltaG}) result from finite, 4--dimensional integrals. Its
general form is
\begin{eqnarray}
&&\Delta_J(r) = \frac23 \bigg[ 2\left(2(n-2)^2-5\right)\Delta_1(r)
+ \left(4(n-2)^2-4\eta(n-2)-5\right)\Delta_2(r)
\nonumber\\
&&\quad{} - 2\left(2(n-2)^2-4\eta(n-2)+9\right)\Delta_3(r) \bigg]\,.
\label{Delta}
\end{eqnarray}
The heavy--flavour contribution, with $m_i=m$, is proportional to
$a_f+\Delta_J(1)$, where
\begin{equation}
\Delta_J(1) = \frac23\left[2(7+2\zeta(2))(n-2)^2
+4(-5+2\zeta(2))\eta(n-2)+17-38\zeta(2)\right]
\label{Delta1}
\end{equation}
is always opposite in sign to $a_f$ and close to it in magnitude, resulting
in a heavy--flavour contribution that is always small, in comparison with
that from a light flavour.

\begin{table}[ht]
\caption{Matching coefficients}
\label{Tab}
\begin{center}
\setlength{\tabcolsep}{0pt}
\begin{tabular}{|c|cccccc|}
\hline
$\Gamma$ & \multicolumn{6}{c|}{$C_\Gamma(m)$} \\
\hline
\hspace{2mm} $1$ \hspace{2mm} &
\hspace{2mm} 1 & $+$ & $\frac23\frac{\alpha_{\rm s}(m)}\pi$ &
$+$ & $(10.92-0.60N_l+0.04)$ &
$\left(\frac{\alpha_{\rm s}}{\pi}\right)^2$ \hspace{2mm} \\[3pt]
\hspace{2mm} $\gamma_0$ \hspace{2mm} &
\hspace{2mm} 1 & $-$ & $\frac23\frac{\alpha_{\rm s}(m)}\pi$ &
$-$ & $(\,4.20-0.44N_l+0.07)$ &
$\left(\frac{\alpha_{\rm s}}{\pi}\right)^2$ \hspace{2mm} \\[3pt]
\hspace{2mm} $\gamma_1$ \hspace{2mm} &
\hspace{2mm} 1 & $-$ & $\frac43\frac{\alpha_{\rm s}(m)}\pi$ &
$-$ & $(11.50-0.79N_l+0.09)$ &
$\left(\frac{\alpha_{\rm s}}{\pi}\right)^2$ \hspace{2mm} \\[3pt]
\hspace{2mm} $\gamma_0\gamma_1$, $\gamma_1\gamma_2$ \hspace{2mm} &
\hspace{2mm} 1 & $-$ & $\frac43\frac{\alpha_{\rm s}(m)}\pi$ &
$-$ & $(16.19-1.13N_l+0.13)$ &
$\left(\frac{\alpha_{\rm s}}{\pi}\right)^2$ \hspace{2mm} \\[3pt]
\hspace{2mm} $\gamma_0\gamma_1\gamma_2$ \hspace{2mm} &
\hspace{2mm} 1 & & &
$-$ & $(10.98-0.77N_l+0.11)$ &
$\left(\frac{\alpha_{\rm s}}{\pi}\right)^2$ \hspace{2mm} \\[3pt]
\hspace{2mm} $\gamma_1\gamma_2\gamma_3$ \hspace{2mm} &
\hspace{2mm} 1 & $+$ & $\frac23\frac{\alpha_{\rm s}(m)}\pi$ &
$-$ & $(\,2.78-0.42N_l+0.08)$ &
$\left(\frac{\alpha_{\rm s}}{\pi}\right)^2$ \hspace{2mm} \\[3pt]
\hspace{2mm} $\gamma_0\gamma_1\gamma_2\gamma_3$ \hspace{2mm} &
\hspace{2mm} 1 & $+$ & $\frac{10}3\frac{\alpha_{\rm s}(m)}\pi$ &
$+$ & $(19.75-0.64N_l+0.00)$ &
$\left(\frac{\alpha_{\rm s}}{\pi}\right)^2$ \hspace{2mm} \\[3pt]
\hline
\end{tabular}
\end{center}
\end{table}

There are 8 distinct matrices, $\Gamma$, in 4--dimensional space--time:
antisymmetrized products of up to 3 spatial $\gamma$--matrices, and the
same products multiplied by $\gamma_0$. For each current, we obtain a
matching coefficient $C_\Gamma(m)$, by choosing appropriate values of $n$
and $\eta$ in the formul\ae~(\ref{cm}) to~(\ref{Delta1}). With $N_l$
zero--mass quarks, and QCD colour factors, we obtain the numerical values
of Table~\ref{Tab}, two of which coincide (for reasons explained in the
next section). Typical matrices $\Gamma$ are presented in the first column;
$C_\Gamma(m)$ is, of course, the same for other spatial components.
Coefficients of $(\alpha_{\rm s}/\pi)^2$ in the two--loop corrections are
written as sums of three terms: the quenched contribution, without quark
loops; the light--quark loop contribution; and the heavy--quark loop
contribution. One can see that the quenched contribution is usually of
order 10; each light flavour gives a contribution of order 1, with the
opposite sign; and the external flavour contributes of order 0.1. In most
cases, the two--loop correction has the same sign as the one--loop
correction.

\section{Ratios of matching coefficients}
\label{ratio} \setcounter{equation}{0}

First we consider ratios that reflect differences in $\gamma_5$
prescriptions. Then we analyze ratios of meson matrix elements.

\subsection{'t~Hooft--Veltman versus anticommuting $\gamma_5$}
\label{Gamma5}

It is generally accepted that there exists no $d$\/--dimensional
generalization of $\gamma_5$ that anticommutes with all $\gamma_\mu$. Nor
is there any covariant $d$\/--dimensional generalization of the
Levi--Civita tensor, $\varepsilon_{\mu\nu\sigma\rho}$. Instead, one may use
the 't~Hooft--Veltman $\gamma_5$,
\begin{equation}
\gamma_5^{\rm HV} = {\rm i}\,\gamma^0\gamma^1\gamma^2\gamma^3\,,
\label{tHV}
\end{equation}
which makes multi--loop calculation rather time--consuming, even when
implemented covariantly, using antisymmetrized
products~\cite{Lar,LV,GL,BK}. However, there is also a general belief that
one may use a naively anticommuting matrix, $\gamma_5^{\rm AC}$, in open
fermion lines, and in loops with an even number of $\gamma_5$--matrices,
without encountering contradictions.

The antisymmetrized products of $n$ and $4-n$ $\gamma$--matrices are
related to each other by multiplication by $\gamma_5^{\rm HV}$, and have
different matching coefficients. On the other hand, multiplying $\Gamma$ by
the naively anticommuting $\gamma_5^{\rm AC}$ does not change $h$
in~(\ref{cond}), and hence cannot change the matching coefficients.
Therefore ratios of matching coefficients with $n$ and $4-n$
antisymmetrized $\gamma$--matrices express differences between using the
different $\gamma_5$ prescriptions.

HQET matrix elements do not depend on the $\gamma_5$ prescription. For
example,
\begin{equation}
{<}0|\bar{q}\gamma_5^{\rm HV}\tilde{Q}|{\rm B}{>} =
{<}0|\bar{q}\gamma_5^{\rm AC}\tilde{Q}|{\rm B}{>}\,.
\label{HQETg5}
\end{equation}
To prove this equality, consider the HQET operator--product
expansion~\cite{BG2} for the correlator of two pseudoscalar currents. In
every contributing diagram, the static--quark propagator between the two
$\gamma_5$--vertices has the $\gamma$--matrix structure $(1+\gamma_0)/2$,
which is unaffected by gluon vertices. Since every sort of
$\gamma_5$--matrix anticommutes with $\gamma_0$, we can move one
$\gamma_5$--matrix from its vertex and annihilate it with the other,
leaving $(1-\gamma_0)/2$ on the heavy--quark line, independently of the
prescription. Since the correlator is independent of the prescription, so
is its spectral density, and in particular the ground--state contribution.
Hence we arrive at~(\ref{HQETg5}).

In contrast to this, the QCD matrix elements do {\em not\/} coincide. We
have already remarked that the $\overline{\rm MS}$--renormalized
pseudoscalar currents, $J^{\rm HV}(\mu) =Z^{-1}_{\rm
HV}(\mu)\bar{q}_0\gamma_5^{\rm HV}Q_0$ and $J^{\rm AC}(\mu)=Z^{-1}_{\rm
AC}(\mu)\bar{q}_0\gamma_5^{\rm AC}Q_0$, have anomalous dimensions that
differ, at two loops and beyond. The currents are related to each other by
a finite renormalization, $J^{\rm AC}(\mu)=Z_P(\mu)J^{\rm HV}(\mu)$, whose
term of order $\alpha_{\rm s}^L$ may be found either by comparing
renormalized matrix elements, at $L$ loops, or more demandingly, by
equating ${\rm d}\log Z_P/{\rm d}\log\mu$ to the difference,
$\gamma_{J^{\rm HV}}-\gamma_{J^{\rm AC}}$, of anomalous dimensions, at
$L+1$ loops. The consistency of these methods has been demonstrated, at
$L=2$, by evaluation of massless three--loop diagrams~\cite{Lar}, which
yield
\begin{equation}
Z_P(\mu) = 1 - 2 C_F \frac{\alpha_{\rm s}(\mu)}{\pi}
+ C_F \left(\frac{\alpha_{\rm s}}{4\pi}\right)^2
\frac{2(C_A+4T_F N_f)}{9}
+{\rm O}(\alpha_{\rm s}^3)\,,
\label{Zp}
\end{equation}
using either method, with a further ${\rm O}(\alpha_{\rm s}^3)$ term
determined solely by the first method. Consistency demands that this result
be obtainable by comparing matrix elements with quarks of any mass and
momenta, and in particular that
\begin{equation}
Z_P(\mu)=
\frac{{<}0|J^{\rm AC}(\mu)|{\rm B}{>}}{{<}0|J^{\rm HV}(\mu)|{\rm B}{>}}
=\frac{C_1(\mu)}{C_{\gamma_0\gamma_1\gamma_2\gamma_3}(\mu)}\,,
\label{CCp}
\end{equation}
given the equality~(\ref{HQETg5}). Setting $\mu=m$ and using~(\ref{cm}), we
indeed verify~(\ref{Zp}). In particular, the two--loop finite--mass
corrections~(\ref{Delta}) cancel in the ratio~(\ref{CCp}), as should occur
in a result obtainable~\cite{Lar} from three--loop mass--independent
anomalous dimensions.

Similarly, the $\overline{\rm MS}$--renormalized 't~Hooft--Veltman and
naively anticommuting non--singlet axial currents are related by a finite
renormalization, $J^{\rm AC}_\mu(\mu)=Z_A(\mu)J^{\rm HV}_\mu(\mu)$, which
has been computed at the one--~\cite{Tru}, two--~\cite{GL}, and
three--loop~\cite{Lar,LV} levels. We are informed by S.~A.~Larin that the
discrepancy between~\cite{Lar,LV} and~\cite{GL} at two loops is
attributable to the use of the so--called G--scheme in~\cite{GL}. The
$\overline{\rm MS}$ result is~\cite{Lar,LV}
\begin{equation}
Z_A(\mu) = 1 - C_F \frac{\alpha_{\rm s}(\mu)}{\pi}
+ C_F \left(\frac{\alpha_{\rm s}}{4\pi}\right)^2
\frac{198C_F-107C_A+4T_F N_f}{9}
+{\rm O}(\alpha_{\rm s}^3)\,,
\label{Za}
\end{equation}
which we verified by calculation of {\em two\/} ratios
\begin{equation}
Z_A(\mu)=\frac{C_{\gamma_0}(\mu)}{C_{\gamma_1\gamma_2\gamma_3}(\mu)}=
\frac{C_{\gamma_3}(\mu)}{C_{\gamma_0\gamma_1\gamma_2}(\mu)}\,.
\label{CCa}
\end{equation}
Note that the weak axial current is $J^{\rm AC}_\mu$. Only for this current
does QCD renormalization preserve a $V-A$ structure. Thus measurable matrix
elements, such as $f_{\rm B}$, are obtained from $J^{\rm AC}_\mu$, not
from the chiral--symmetry--breaking 't~Hooft--Veltman counterpart.

The QCD tensor current, $J_{\mu\nu}=\bar{q}\sigma_{\mu\nu}q$, is a special
case, because inclusion of $\gamma_5^{\rm HV}$, to form $J_{\mu\nu}^{\rm
HV}=\bar{q}\gamma_5^{\rm HV}\sigma_{\mu\nu}q$, is merely a space--time
transformation, in 4 dimensions, giving, for example, $J_{01}^{\rm
HV}=-{\rm i}J_{23}$. Thus there can be no difference of anomalous
dimensions between $J_{\mu\nu}^{\rm HV}$ and $J_{\mu\nu}^{\rm AC}$, and
hence no non--trivial finite renormalization, $Z_T(\mu)$, relating the two
prescriptions. Equivalently, we have {\em two\/} methods of obtaining
$Z_T(\mu)$:
\begin{equation}
Z_T(\mu)=
\frac{C_{\gamma_0\gamma_1}(\mu)}{C_{\gamma_2\gamma_3}(\mu)}=
\frac{C_{\gamma_2\gamma_3}(\mu)}{C_{\gamma_0\gamma_1}(\mu)}\,,
\label{CCt}
\end{equation}
which proves both that $Z_T(\mu)=1$ and that
$C_{\gamma_0\gamma_1}(m)=C_{\gamma_2\gamma_3}(m)$, as found
from~(\ref{cm}). It is easy to see how this equality comes about, purely
within our calculation of matching coefficients. With $n=2$, we have
$h=\eta\varepsilon$, so that the only possible origin of a difference
between $C_{\gamma_0\gamma_1}(\mu)$ and $C_{\gamma_2\gamma_3}(\mu)$ would
be a term $h/\varepsilon$ in $\Gamma_0$. Such a term would give an
$\eta$--dependent QCD anomalous dimension, for $n\ne2$, and hence cannot
occur.

\subsection{Ratios of meson matrix elements}
\label{Meson}

After eliminating the currents containing the 't~Hooft--Veltman $\gamma_5$,
we are left with 4 essentially different currents. In this subsection, we
discuss the ground--state--meson matrix elements:
\begin{eqnarray}
&&{<}0|(\bar{q}\gamma_5 Q)_\mu|{\rm B}{>}
=-{\rm i}\,m_{\rm B} f_{\rm B}^P(\mu)\,,
\nonumber\\
&&{<}0|\bar{q}\gamma_\alpha\gamma_5 Q|{\rm B}{>}
={\rm i}\,f_{\rm B}p_\alpha\,,
\label{mme}\\
&&{<}0|\bar{q}\gamma_\alpha Q|{\rm B}^*{>}
={\rm i}\,m_{{\rm B}^*} f_{{\rm B}^*} e_\alpha\,,
\nonumber\\
&&{<}0|(\bar{q}\sigma_{\alpha\beta}Q)_\mu|{\rm B}^*{>}
=f_{{\rm B}^*}^T(\mu) (p_\alpha e_\beta-p_\beta e_\alpha)\,,
\nonumber
\end{eqnarray}
where $\sigma_{\mu\nu}=\frac{\rm i}2[\gamma_\mu,\gamma_\nu]$, and we use,
from now on, the naively anticommuting $\gamma_5$ that is appropriate to
matrix elements of non--singlet weak currents. The corresponding HQET
quantities coincide:
$\tilde{f}_{\rm B}^P = \tilde{f}_{\rm B} = \tilde{f}_{{\rm B}^*}
= \tilde{f}_{{\rm B}^*}^T$, as consequences of the heavy--quark symmetry.
Hence, ratios of QCD matrix elements are equal to ratios of matching
coefficients. These ratios do not depend on the unknown three--loop HQET
anomalous dimension. Similar formul{\ae} hold for P--wave $0^+$, $1^+$
mesons, if one inserts an extra $\gamma_5$ into all currents.

It follows from the equations of motion that
\begin{equation}
\frac{f^P_{\rm B}(\mu)}{f_{\rm B}}=
\frac{{<}0|(\bar{q}\gamma_5 Q)_\mu|{\rm B}{>}}
{{<}0|\bar{q}\gamma_5\gamma_0 Q|{\rm B}{>}}
= \frac{m_{\rm B}}{\overline{m}(\mu)}\,,
\label{mm}
\end{equation}
where $\overline{m}(\mu)$ is the $\overline{\rm MS}$ running heavy--quark
mass. To leading order in $1/m$, we may replace $m_{\rm B}$ by the
on--shell mass $m$, obtaining
\begin{eqnarray}
&&\frac{f^P_{\rm B}(m)}{f_{\rm B}}=
\frac{m}{\overline{m}(m)}=
\frac{C_1(m)}{C_{\gamma_0}(m)}
= 1 + C_F \frac{\alpha_{\rm s}(m)}{\pi}
\nonumber\\
&&\quad C_F \left(\frac{\alpha_{\rm s}}{4\pi}\right)^2
\bigg[ C_F \left(\frac{121}{8}+30\zeta(2)-8I\right)
+ C_A \left(\frac{1111}{24}-8\zeta(2)+4I\right)
\nonumber\\
&&\quad{}
+ 8 T_F \sum_{i=1}^{N_f} \left\{-\frac{71}{48}-\zeta(2)
+\Delta_1\left(\frac{m_i}{m}\right)
+\Delta_3\left(\frac{m_i}{m}\right)\right\}\bigg]
\label{Rm}\\
&&\quad \approx 1 + \frac43 \frac{\alpha_{\rm s}(m)}{\pi}
+ (16.01-1.04N_l+0.10) \left(\frac{\alpha_{\rm s}}{\pi}\right)^2\,,
\nonumber
\end{eqnarray}
where the final numerical result shows, as in Table~\ref{Tab}, the
contributions of: the quenched term; $N_l$ massless--quark loops; and the
heavy--quark loop. We confirm the analytical result of~\cite{GBGS}, in the
case of $N_l=N_f-1$ massless quarks, and correct the omission of a factor
$C_F=4/3$ from the numerically much smaller effects of finite light--quark
masses in Equation~(17) of~\cite{GBGS}.

Our most important result is for the ratio of two observable matrix
elements:
\begin{equation}
\frac{{<}0|\bar{q}\vec{\gamma}Q|{\rm B}^*{>}}
{{<}0|\bar{q}\gamma_0\gamma_5 Q|{\rm B}{>}}
=\frac{m_{{\rm B}^*}f_{{\rm B}^*}\vec{e}}{m_{\rm B}f_{\rm B}}\,,
\label{ff}
\end{equation}
which is independent of the renormalization scale. At leading order in
$1/m$, we obtain
\begin{eqnarray}
&&\frac{f_{{\rm B}^*}}{f_{\rm B}}
= \frac{C_{\gamma_1}(m)}{C_{\gamma_0}(m)}
= 1 - C_F \frac{\alpha_{\rm s}(m)}{2\pi}
\nonumber\\
&&\quad{} + C_F \left(\frac{\alpha_{\rm s}}{4\pi}\right)^2
\bigg[ C_F \frac13(31-128\zeta(2)+16I)
+ C_A \frac19(-263+144\zeta(2)-24I)
\nonumber\\
&&\quad{}
- \frac{16}3 T_F \sum_{i=1}^{N_f} \left\{
-\frac{19}{12}
+\Delta_2\left(\frac{m_i}{m}\right)
-2\Delta_3\left(\frac{m_i}{m}\right)\right\}\bigg]
\label{Rvp}\\
&&\quad \approx 1 - \frac23 \frac{\alpha_{\rm s}}{\pi}
- (7.75-0.35N_l+0.03) \left(\frac{\alpha_{\rm s}}{\pi}\right)^2.
\nonumber
\end{eqnarray}
Note that this is numerically very close to $(\overline{m}(m)/m)^{1/2}$, as
can be seen from~(\ref{Rm}).

At any finite order of perturbation theory, the ratio~(\ref{Rvp}) is
determined by a combination of loop integrals that contains neither
infrared nor ultraviolet divergences. Therefore all loop momenta are of
order $m$. In order to demonstrate the absence of infrared sensitivity
(which may sometimes be overlooked in dimensional regularization) we
repeated the one--loop calculation with a gluon mass, $\lambda$, finding
that the only effect is to multiply the ${\rm O}(\alpha_{\rm s})$ term
in~(\ref{Rvp}) by
\begin{eqnarray}
&&G\left(\frac{\lambda}{m}\right) = \frac13 \int\limits_0^\infty
\frac{{\rm d}k^2}{k^2+\lambda^2} F\left(\frac{k^2}{m^2}\right)
= 1 - \frac{2\pi}{3} \frac{\lambda}{m}
+ {\rm O}\left(\frac{\lambda^2}{m^2}\right)\,,
\label{glum}\\
&&F(x) = (x+1)\sqrt{x(x+4)}-x(x+3) =
\left\{
\begin{array}{ll}
2\sqrt{x}, & x\ll1\,, \\
2/x, & x\gg1\,.
\end{array}
\right.
\nonumber
\end{eqnarray}
In general, we find that all ratios of matching coefficients are
infrared--safe, though those involving QCD currents with anomalous
dimensions clearly require ultraviolet regularization.

We now demonstrate that the existence of a very heavy flavour, such as top,
has {\em no\/} effect on the observable ratio $f_{{\rm B}^*}/f_{\rm B}$. To
prove this decoupling theorem, we rewrite~(\ref{Rvp}) as
\begin{eqnarray}
&&\frac{f_{{\rm B}^*}}{f_{\rm B}}
= 1 - C_F \frac{\alpha_{\rm s}^{[N_f]}(\mu)}{2\pi}
+ C_F \left(\frac{\alpha_{\rm s}}{4\pi}\right)^2
\left[A + B^{[N_f]}(\mu)\right]\,,
\label{deco}\\
&&B^{[N_f]}(\mu) = \frac{44}{3} C_A \log\frac{m}{\mu}
- \frac{16}{3} T_F \sum_{i=1}^{N_f}
\left\{\log\frac{m}{\mu} - \frac{19}{12}
+ \Delta_2\left(\frac{m_i}{m}\right)
- 2\Delta_3\left(\frac{m_i}{m}\right)\right\}\,,
\nonumber
\end{eqnarray}
where $A$ is a constant, specifying the quenched term in~(\ref{Rvp}), and
we have transformed the $\overline{\rm MS}$ coupling to an arbitrary scale
$\mu$, using the one--loop $N_f$--flavour $\beta$--function. The explicit
$\mu$--dependence in~(\ref{deco}) at ${\rm O}(\alpha_{\rm s}^2)$ is
cancelled by the implicit dependence at ${\rm O}(\alpha_{\rm s})$. Now,
suppose that we take account of the existence of a super--heavy quark, with
mass $m_h\gg m$. The effect is merely to replace $N_f$ by $N_f+1$
in~(\ref{deco}), where
\begin{equation}
\frac{\alpha_{\rm s}^{[N_f+1]}(\mu)}{2\pi}
=\frac{\alpha_{\rm s}^{[N_f ]}(\mu)}{2\pi}
-\frac{16}{3}T_F\left(\frac{\alpha_{\rm s}}{4\pi}\right)^2
\left\{\log\frac{m_h}{\mu}\right\}
\label{imp}
\end{equation}
ensures that $\alpha_{\rm s}^{[N_f+1]}(m_h)=\alpha_{\rm s}^{[N_f]}(m_h)$,
and
\begin{equation}
B^{[N_f+1]}(\mu)
=B^{[N_f ]}(\mu)
-\frac{16}{3}T_F\left\{
\log\frac{m}{\mu} - \frac{19}{12}
+\Delta_2\left(\frac{m_h}{m}\right)
-2\Delta_3\left(\frac{m_h}{m}\right)\right\}
\label{exp}
\end{equation}
includes the loop--effect of the super--heavy quark in diagram~\ref{Fig}b.
Referring to~(\ref{larger}), we find that $\Delta_2(r)-2\Delta_3(r)=\log
r+\frac{19}{12}+{\rm O}(r^{-2}\log r)$, for $r=m_h/m\gg1$. Hence the terms
in braces in~(\ref{imp}) and~(\ref{exp}) cancel, at any scale $\mu$. This
also demonstrates that one must match the $\overline{\rm MS}$ couplings at
$\mu=m_h$, as in~(\ref{imp}), and not at, say, $\mu=2m_h$. Interestingly,
the external heavy flavour is also numerically unimportant. Its
contribution to~(\ref{Rvp}), relative to that of a massless quark, is
suppressed by a factor
\begin{equation}
1-{\textstyle\frac{12}{19}}\left\{\Delta_2(1)-2\Delta_3(1)\right\}
={\textstyle\frac1{19}}(4\pi^2-41)=-0.08\,.
\label{approx}
\end{equation}
Comparable suppressions occur in all our results.

Turning to the numerical significance of our two--loop result, we
substitute $N_l=4$ in~(\ref{Rvp}), and add the $1/m$ correction
of~\cite{Neu2,Ball}, obtaining
\begin{equation}
\frac{f_{{\rm B}^*}}{f_{\rm B}} = 1 - \frac23 \frac{\alpha_{\rm s}(m_b)}{\pi}
- \left(\frac{\alpha_{\rm s}}{\pi}\right)^2
\left(6.37+\Delta_c\right)
+ \frac{1}{m_b} \left(\frac23\bar{\Lambda}-8G_2(m_b)\right)
+{\rm O}(\alpha_{\rm s}^3,\alpha_{\rm s}/m_b,1/m_b^2)\,,
\label{fbnum}
\end{equation}
with a finite $c$--quark mass correction
$\Delta_c=\frac29(\Delta_2(m_c/m_b)-2\Delta_3(m_c/m_b))=0.18\pm0.01$, for a
mass ratio $m_c/m_b=0.28\pm0.03$, obtained from $m_b=(4.8\pm0.2)$~GeV,
using $m_{\rm B,D}=m_{b,c}+\bar{\Lambda}+(\mu_\pi^2-\mu_G^2)/(2m_{b,c})$
and $m_{{\rm B}^*}=m_{\rm B}+2\mu_G^2/(3m_b)$ with a kinetic term
$\mu_\pi^2=(0.5\pm0.1)$~GeV$^2$. The current world average $\alpha_{\rm
s}(m_Z)=0.117\pm0.005$~\cite{PDG}, evolved down to $m_b$ using the
three--loop formula, gives $\alpha_{\rm s}(m_b)=0.215\pm0.018$. Therefore
the one--loop correction is $-(4.6\pm0.4)\%$, and the two--loop term is
comparable to it: $-(3.1\pm0.5)\%$. Another way to state the slow
convergence of the perturbation series is to say that the
fastest--apparent--convergence scheme, with a vanishing two--loop
correction in~(\ref{deco}), would require one to evaluate $\alpha_{\rm
s}(\mu)$ at far too low a scale: $\mu=m_b/13=370$~MeV.

The chromomagnetic interaction matrix element $G_2(m_b)$ is not well known;
sum--rule estimates range between $G_2(m_b)=-26$~MeV~\cite{Neu2}, and
$G_2(m_b)=+21$~MeV~\cite{Ball}, with large uncertainties. Including the
meson residual energy, $\bar{\Lambda}$, one obtains a $1/m$ correction of
$+(11\pm3)\%$, from~\cite{Neu2}, and $+(4\pm3\pm2\pm2)\%$,
from~\cite{Ball}, whose three sources of uncertainty arise from: sum--rule
fitting; an unknown two--loop anomalous dimension; and an uncertainty in
$\bar{\Lambda}$. Combining our two--loop radiative corrections with this
range of $1/m$ effects, we arrive at $f_{{\rm B}^*}/f_{\rm B}=1.00\pm0.04$,
with an uncertainty reflecting the difference between~\cite{Neu2}
and~\cite{Ball}. It is, however, quite unclear whether the perturbation
series in $\alpha_{\rm s}$ is converging fast enough for us to neglect the
unknown three--loop term.

For the last ratio,
\begin{equation}
\frac{{<}0|(\bar{q}\gamma_i\gamma_0 Q)_\mu|{\rm B}{>}}
{{<}0|\bar{q}\gamma_i Q|{\rm B}{>}}
= \frac{f_{{\rm B}^*}^T(\mu)}{f_{{\rm B}^*}}\,,
\label{ft}
\end{equation}
we obtain
\begin{eqnarray}
&&\frac{f_{{\rm B}^*}^T(m)}{f_{{\rm B}^*}}
= \frac{C_{\gamma_0\gamma_1}(m)}{C_{\gamma_1}(m)} = 1
\nonumber\\
&&\quad{} + C_F \left(\frac{\alpha_{\rm s}}{4\pi}\right)^2
\bigg[ C_F \frac13\left(\frac{307}{8}-70\zeta(2)+8I\right)
+ C_A \frac13\left(-\frac{4277}{72}+24\zeta(2)-4I\right)
\nonumber\\
&&\quad{}
- \frac83 T_F \sum_{i=1}^{N_f} \left\{
-\frac{205}{144}-\zeta(2)
+\Delta_1\left(\frac{m_i}{m}\right)
+\Delta_3\left(\frac{m_i}{m}\right)\right\}\bigg]
\label{Rtv}\\
&&\quad \approx 1 - (4.69-0.34N_l+0.04)
\left(\frac{\alpha_{\rm s}}{\pi}\right)^2\,.
\nonumber
\end{eqnarray}
The one--loop term vanishes at $\mu=m$, but will of course appear at any
other $\mu$. If one wants $f^T_{{\rm B}^*}(\mu)$ at a scale $\mu$ widely
separated from $m$, one needs to use the QCD three--loop anomalous
dimension of the current $\bar{q}\sigma_{\mu\nu}q$, in conjunction
with~(\ref{Rtv}). This anomalous dimension is not known at present, though
it could be calculated using standard massless--quark methods.

Finally we note an intriguing pattern in the
results~(\ref{Rm}),~(\ref{Rvp}) and~(\ref{Rtv}): in each a reasonable
estimate of the two--loop term is obtained by the simple device of
replacing $N_l$ by $N_l-\frac{33}{2}$ in the easily computed
light--quark--loop contributions. We call this device ``naive
non--abelianization'', since it is based on the hope that results in the
non--abelian theory may be estimated merely by replacing the leading term
in the abelian large--$N_f$ $\beta$\/--function by its non--abelian
counterpart. Comparing naively non--abelianized estimates with our exact
two--loop coefficients, we find that, with $N_l=4$, the two--loop term in
$m/\overline{m}(m)$ is overestimated by only 9\% in~(\ref{Rm}), whilst
in~(\ref{Rvp}) one obtains an underestimate by 31\%, and in~(\ref{Rtv}) an
overestimate by 27\%. We thus see some merit in estimating radiative
corrections by naive non--abelianization, in cases where a large--$N_f$
calculation is practicable, whilst an exact one is not. It is noteworthy
that naive non--abelianization of the large--$N_f$ terms~\cite{BK} in
${\rm e}^+{\rm e}^-$ annihilation underestimates the three--loop terms in
$R$ by less than 6\%, with 4 or 5 active quark flavours.

\section{All--order results}
\label{higher} \setcounter{equation}{0}

We now apply the methods of~\cite{PMP,LNF,Ben} to study the matching
coefficients, at all orders of perturbation theory, in a highly fictitious
limit: $N_f\to-\infty$.

As in other recent studies~\cite{BB,Bigi,NS}, the intention is to
investigate so--called renormalon effects, associated with factorial growth
of the coefficients of perturbation series~\cite{Mue}. Unfortunately, there
is no gauge--invariant prescription that enables one to obtain all--order
results in the asymptotically--free theory of interest; instead one merely
studies a single chain of fermion loops, such as that which gives rise to
the Landau pole in QED. The hope is that one may learn something
about the non--abelian theory, by imagining that asymptotic freedom still
holds with an infinite number of massless flavours. In practice, the
analysis amounts to no more than changing the sign of $N_f$ in the
large--$N_f$ methods of~\cite{PMP,LNF,Ben}.

\subsection{Master formula}

Following recent practice, we disguise the sleight--of--hand, by hiding the
large--$N_f$ contribution to the $\beta$--function in a (fictitiously)
positive value of $\beta_0=\frac13(11C_A-4T_F N_f)$. In other words, we
imagine that $\beta=-{\rm d}\log\alpha_{\rm s}/{\rm d}\log\mu=
\beta_0\alpha_{\rm s}/2\pi+ {\rm O}(\alpha_{\rm s}^2)$ remains finite and
positive, so that the QCD behaviour $\beta\sim1/\log(\mu/\LM)$ still holds
at large $\mu$.

In the matching coefficient we retain only the leading terms, of order
$1/\beta_0$, in the limit $\beta_0\to\infty$, whilst retaining all powers
of $\beta\sim1/\log(\mu/\LM)$. As the heavy--quark loop is negligible, in
comparison with loops from a large number of massless quarks, we
immediately obtain $\tilde{\Gamma}_0=\Gamma_{\rm B}$ and $\tilde{Z}_Q^{\rm
os}=1$. A single $L$\/--loop diagram contributes to $\Gamma_0$. It is
obtained by inserting a chain of $L-1$ massless--quark loops in the gluon
line of diagram~\ref{Fig}a. A corresponding insertion yields the
$L$\/--loop diagram contributing to $Z_Q^{\rm os}$. As $\beta_0\to\infty$,
the multiplications in~(\ref{main}) degenerate to mere addition and
subtraction of terms of order $1/\beta_0$. Hence multiplicative
renormalization of the QCD and HQET currents amounts to no more than
minimal subtraction of powers of $1/\varepsilon$. Moreover, there is no
mass counterterm to consider at ${\rm O}(1/\beta_0)$, and
coupling--constant renormalization amounts only to
$\beta_0g_0^2/(4\pi)^2=\bar\mu^{2\varepsilon}\beta/(2+\beta/\varepsilon)$,
with $\bar\mu^2=\mu^2{\rm e}^\gamma/4\pi$. Hence the entire perturbation
series may be written, formally, as follows
\begin{equation}
C_\Gamma(\mu)=1+\sum_{L=1}^\infty\frac{F(\varepsilon,L\varepsilon)}{L}
\left(\frac{\beta}{2\varepsilon+\beta}\right)^L
-(\mbox{minimal subtractions})
+{\rm O}\left(\frac{1}{\beta_0^2}\right)\,,
\label{formal}
\end{equation}
where $F(\varepsilon,u)$ is regular at $\varepsilon=u=0$, and is easily
calculated, in terms of a massless one--loop integral, in the gluon proper
self--energy, and two massive on--shell one--loop integrals, determining
the contributions of $Z_Q^{\rm os}$ and $\Gamma_0$.

As in the case of deep--inelastic sum rules at large $N_f$~\cite{BK}, the
computational burden is slight, in comparison with the analysis of $F_{32}$
hypergeometric functions in the two--loop integrals required for current
correlators in QED~\cite{LNF} and HQET~\cite{BB}. We find that
\begin{eqnarray}
&&\hspace{-12mm}
F(\varepsilon,u)=-\frac{C_F}{\beta_0}
\left(\frac{\mu}{m}\right)^{2u}
\frac{\Gamma(1+u)\Gamma(1-2u)}
{{\rm e}^{-\gamma\varepsilon}\Gamma(3-u-\varepsilon)}
\frac{N(\varepsilon,u)}
{\left[D(\varepsilon)\right]^{1-u/\varepsilon}}\,,
\label{Fed}\\
&&\hspace{-12mm}
N(\varepsilon,u)=(3-2\varepsilon)(1-u)(1+u-\varepsilon)+2
-u-\varepsilon+2\eta(n-2+\varepsilon)u-2(n-2+\varepsilon)^2,\,{}
\label{Ned}\\
&&\hspace{-12mm}
D(\varepsilon)=6{\rm e}^{\gamma\varepsilon}
\Gamma(1+\varepsilon)B(2-\varepsilon,2-\varepsilon)
=1+{\textstyle\frac53}\varepsilon+{\rm O}(\varepsilon^2)\,.
\label{De}
\end{eqnarray}
The first term in~(\ref{Ned}) derives from $Z_Q^{\rm os}$; the
remaining terms, from $\Gamma_0$, are quadratic in
$h=\eta(n-2+\varepsilon)$, since they result from an essentially
one--loop calculation.
The function~(\ref{De}) comes from the gluon self--energy.

We now follow the methods of~\cite{PMP,LNF}, expanding
$F(\varepsilon,L\varepsilon)$ in powers of $\varepsilon$ and
$L\varepsilon$, and expanding $\left[\beta/(2\varepsilon+\beta)\right]^L$
in powers of $\beta/2\varepsilon$, to obtain a quadruple sum
in~(\ref{formal}). As shown in~\cite{PMP}, combinatoric identities relate
$1/\varepsilon$ terms, and hence $\overline{\rm MS}$ subtractions, to the
Taylor coefficients of $F(\varepsilon,0)$. In the case of the matching
coefficient~(\ref{main}), we simply obtain
\begin{equation}
\tilde{\gamma}_J-\gamma_J=\beta F(-\beta/2,0)
+{\rm O}\left(\frac{1}{\beta_0^2}\right)\,.
\label{Fb}
\end{equation}
As shown in~\cite{LNF}, the finite terms receive contributions from the
Taylor coefficients of $F(\varepsilon,0)$ and also from the Taylor
coefficients of $F(0,u)$. The former are scheme--dependent, and give a
well--behaved series in $\beta$. The latter are scheme--independent, and
give a series that is not Borel--summable. A formal statement of the result
may be written very simply:
\begin{equation}
C_\Gamma(\mu)=1+\int\limits_{-\frac{\beta}{2}}^0
{\rm d}\varepsilon\,\frac{F(0,0)-F(\varepsilon,0)}{\varepsilon}
+\int\limits_0^\infty
{\rm d}u\exp\left(\frac{-2u}{\beta}\right)\frac{F(0,u)-F(0,0)}{u}
+{\rm O}\left(\frac{1}{\beta_0^2}\right)\,.
\label{path}
\end{equation}

The function $F(0,u)$ specifies the Borel transform of
the scheme--independent contributions~\cite{BB,Bigi}. Attempting to undo
the Borel transform, via the notional Laplace transform of~(\ref{path}),
one encounters singularities at $u>0$, which are referred to as infrared
renormalons~\cite{Mue}. They reflect the fact that at higher and higher
orders in perturbation theory one is probing regions of smaller and smaller
gluon momenta. The first infrared renormalon occurs at $u=\frac12$,
giving a singularity in the integral with a residue that is a multiple of
$\exp(-1/\beta)\mu/m\sim\LM/m$. Any attempt to sum the perturbation series
involves an arbitrary choice of how to deal with this singularity. In
general, a singularity at $u=u_0>0$ has a residue that is a multiple of
$(\LM/m)^{2u_0}$. Thus the perturbation series, taken to all orders, has
ambiguities that are formally commensurate with the higher--dimension
operators in the $1/m$ expansion~(\ref{match}).

The existence of such ambiguities is profoundly unsurprising. We see
from~(\ref{glum}) that the introduction of an infrared regulator,
$\lambda$, such as a gluon mass, would modify the result of on--shell
integrals at order $\lambda/m$. The infrared renormalon at $u=\frac12$
serves to remind one of this simple fact. The situation is in close analogy
with the operator--product expansion of QCD current correlators, at large
$Q^2=-q^2$, where the first infrared renormalon, at $u=2$, leads to an
ambiguity of order $(\LM/Q)^4$, commensurate with the gluon--condensate
contribution~\cite{Mue}. Thus one should not flinch at renormalons; they
serve as healthy reminders that one has chosen to integrate over all gluon
momenta, including the infrared region where non--perturbative effects are
dominant. In fact, infrared renormalons have a positive virtue: the
pattern of their residues must match the contributions of higher--dimension
operators and thus provides consistency checks on the form of the $1/m$
expansion of HQET, or the $1/Q^2$ expansion of massless QCD. Since we have
not taken the (inordinate) trouble of using an infrared regulator, the
factorization of short-- and long--distance physics into coefficients and
operators is intrinsically ambiguous; each resummation prescription
corresponds to a different set of values for the operator matrix elements.
Nature, however, is not as slip--shod as we; she takes care to ensure that
physical quantities are independent of the exigencies of our calculations.

Hence we see, from~(\ref{Fb}) and~(\ref{path}), that the simple
multinomial~(\ref{Ned}) furnishes detailed information about anomalous
dimensions, through its $\varepsilon$\/--dependence, and about renormalon
singularities, through its $u$\/--dependence. Moreover, it does so for all
current matchings, with $n$ specifying the QCD current and $\eta=\pm1$ the
specific components that are matched to HQET currents. We now unpack some
of this wealth of information.

\subsection{Anomalous dimensions}

Since we know that $\gamma_J$ vanishes at $n=1$, we can obtain both
$\gamma_J$ and $\tilde{\gamma}_J$ from~(\ref{Fb}):
\begin{equation}
\gamma_J=\left.C_F\frac{\alpha_{\rm s}}{2\pi}\frac{(n-1)(d-1-n)}
{18B(d/2,d/2)B(d/2+1,3-d/2)}\right|_{d=4+\beta}\,,\quad
\tilde{\gamma}_J=\left.{\textstyle\frac12}\gamma_J\right|_{n=0}\,,
\label{agj}
\end{equation}
at order $1/\beta_0$. For the QCD currents, we verify the $n=0$ result
of~\cite{PMP}, and the $n=3$ analysis of~\cite{BK}. The $n=2$ and $n=4$
results, as well as that for $\tilde{\gamma}_J$, are, we believe, new.
It is rather intriguing that
$\tilde{\gamma}_J-\frac12\gamma_{\bar{q}q}$ vanishes at ${\rm
O}(1/\beta_0)$, thereby continuing, at all orders, a trend already apparent
in the simple $N_f$--independent two--loop formula given for this
combination in~\cite{BG}. A consequence of the all--order result is that
the coefficient of the light--quark condensate, in the operator--product
expansion of the HQET current correlator, is scale--independent, at leading
order in $1/\beta_0$. This is confirmed, at two loops, by~\cite{BG2}. The
all--order result may be somewhat accidental, since all large--$N_f$
anomalous dimensions are fairly simply related. For example, we find, by
similar methods, that the QCD field, $q$, and the HQET field, $\tilde{Q}$,
have the following large--$N_f$ off--shell anomalous dimensions:
\begin{equation}
\gamma_q=a C_F\frac{\alpha_{\rm s}}{\pi}
-{\textstyle\frac12}\beta\tilde{\gamma}_J\,,\quad
\tilde{\gamma}_Q=a C_F\frac{\alpha_{\rm s}}{\pi}
+{\textstyle\frac12}(4+\beta)\tilde{\gamma}_J\,,
\label{aqq}
\end{equation}
at order $1/\beta_0$. The dependence on the gauge parameter, $a$, is
limited to the one--loop level. In the Landau gauge, with $a=0$,
the light--quark result agrees with~\cite{JAG} and the heavy--quark
result with~\cite{BB}. Impressively, all the ${\rm O}(1/\beta_0^2)$ terms
in the anomalous dimension of the electron field were obtained in the
QED analysis of~\cite{JAG}.

We have a strong check of~(\ref{aqq}). The HQET current anomalous
dimension is
\begin{equation}
\tilde{\gamma}_J = \tilde{\gamma}_\Gamma
+ {\textstyle\frac12} (\gamma_q+\tilde{\gamma}_Q)\,,
\label{aJt}
\end{equation}
where $\tilde{\gamma}_\Gamma={\rm d}\log \tilde{Z}_\Gamma/{\rm d}\log\mu$,
$\tilde{Z}_\Gamma$ being the renormalization constant of the HQET
proper--vertex function. It can be obtained from the $1/\varepsilon$ pole
of the proper vertex at zero light--quark momentum and non--zero heavy--quark
residual energy, because there are no infrared divergences in this case.
At order $1/\beta_0$, all vertex diagrams with more than one loop
contain a transverse gluon propagator, with quark--loop insertions.
The argument in Section~2.1 shows that such diagrams vanish.
Therefore, at ${\rm O}(1/\beta_0)$, there are no multi--loop contributions
to $\tilde{\gamma}_\Gamma$. Moreover, the one--loop contribution vanishes
in the Landau gauge. Taking account of this, the field anomalous
dimensions~(\ref{aqq}) reproduce $\tilde{\gamma}_J$.

One easily obtains the perturbative expansions of~(\ref{agj})
and~(\ref{aqq}), from the $\varepsilon$--expansion
\begin{eqnarray}
&&\hspace{-10mm}\frac{1}{18B(d/2,d/2)B(d/2+1,3-d/2)}
=\frac{\sum_{k=0}^\infty\left(\frac{1}{2^k}-\frac{2k}{3}\right)
\varepsilon^k}{\exp\left\{\sum_{s=3}^\infty\left(3+[-1]^s-2^s\right)
\zeta(s)\varepsilon^s/s\right\}}
\label{epe}\\[3pt]
&&\quad{}=1-{\textstyle\frac16}\varepsilon
-{\textstyle\frac{13}{12}}\varepsilon^2
+\left(2\zeta(3)-{\textstyle\frac{15}{8}}\right)\varepsilon^3
+\left(3\zeta(4)-{\textstyle\frac{125}{48}}\right)\varepsilon^4
+{\rm O}\left(\varepsilon^5\right)\,,
\nonumber
\end{eqnarray}
which reproduces the large--$N_f$\/ features of all existing two-- and
three--loop calculations, and provides checks for future perturbative
calculations. It is notable that $\zeta(4)$, which is conspicuously absent
from three-- and four--loop QCD results, is bound to occur at higher
orders. Whilst the perturbative expansion~(\ref{epe}) is valid only for
$|\varepsilon|<\frac12$, corresponding to $|\beta|<1$, one can see from the
singularities of the Euler--Beta functions in~(\ref{epe}) that in fact no
singularity is encountered for $d>-1$. Hence the anomalous dimensions are
well defined for $\beta>-5$, and summation of the large--$N_f$ perturbation
series results in an extension of the domain of convergence.

\subsection{Renormalon ambiguities and their cancellation}

In stark contrast to the first integral in~(\ref{path}), which exists for
all $\beta>-5$, the second contains infrared renormalons, for all
$\beta>0$. If one arbitrarily chose to evaluate it by principal--value
integration, one would differ, at order $\LM/m$, with someone who chose to
add some (presumably real) multiple of the residue at $u=\frac12$. A
solution would be to absorb this disagreement into different values adopted
for the other terms that occur at ${\rm O}(1/m)$ in~(\ref{match}).

The $1/m$ corrections to the ground--state meson matrix elements were
calculated in~\cite{Neu2} for the currents with $n=1$. We performed the
corresponding calculation for an arbitrary $\Gamma$, obtaining a general
result in terms of $h|_{\varepsilon=0}=\eta(n-2)$:
\begin{eqnarray}
&&\frac{{<}0|J(\mu)|{\rm M}{>}}{{<}0|\tilde{J}(\mu)|{\rm M}{>}}=
C_\Gamma(\mu) + \frac1m\left(G_1+2d_\Gamma G_2
-\frac16\bar{d}_\Gamma(m_{\rm M}-m)\right)
+{\rm O}\left(\frac{1}{m^2}\right)\,,
\label{1/m}\\
&&\bar{d}_\Gamma=1-2\eta(n-2)\,,\quad
d_\Gamma={\textstyle\frac12}\left(\bar{d}_\Gamma^2-3\right)\,,
\nonumber
\end{eqnarray}
where M is the ground--state or excited--state meson that couples
to the current, and $G_1$ and $G_2$ are bilocal matrix elements of the
heavy--quark kinetic energy and chromomagnetic interaction,
respectively.
Multiplying $\Gamma$ by $\gamma_0$ changes the
sign of $\bar{d}_\Gamma$, without changing
$d_\Gamma$, since the latter is obtained from
$\sigma_{\mu\nu}\Gamma\frac{1+\gamma_0}{2}\sigma^{\mu\nu}
\frac{1+\gamma_0}{2}=2d_\Gamma\Gamma\frac{1+\gamma_0}{2}$.
We find that: $\bar{d}_\Gamma=(-1)^{n+1}d_\Gamma$
for all 8 currents; $d_\Gamma=3$ for the 4 currents
that couple to $0^\pm$ mesons; $d_\Gamma=-1$ for those
that couple to $1^\pm$ mesons. The ground--state S--wave $0^-$ and $1^-$
mesons remain degenerate at this order in $1/m$, as do the excited--state
P--wave $0^+$ and $1^+$ mesons.

Viewed from the standpoint of HQET, the residual--energy term,
$\bar{\Lambda}=m_{\rm B}-m_b+{\rm O}(1/m_b)$, and the matrix elements,
$G_{1,2}$, suffer from ultraviolet--renormalon ambiguities~\cite{BB}.
HQET knows nothing about the pole mass, $m_b$, since it deals only with
residual energies. However, the HQET self--energy, $\tilde\Sigma(\omega)$,
has a linear ultraviolet divergence, conventionally suppressed by
dimensional regularization. If one were to use an ultraviolet
momentum--space cut--off, it would be necessary to introduce a
residual mass, into which this linear divergence could be absorbed.
Moreover, $G_{1,2}$ are bound to acquire ultraviolet--renormalon ambiguities,
via mixing with $\bar{\Lambda}$~\cite{NS}.
We now determine these by demanding that they cancel
the infrared--renormalon ambiguities in~(\ref{1/m}).

Setting $\varepsilon=0$ in the master result~(\ref{Fed}) and~(\ref{Ned}),
we obtain
\begin{equation}
F(0,u)=-\frac{C_F}{\beta_0}
\left(\frac{\mu{\rm e}^{\frac56}}{m}\right)^{2u}
\frac{B(1+u,1-2u)}{2-u}
\left[5-u-3u^2+2\eta(n-2)u-2(n-2)^2\right]\,.
\label{Fu}
\end{equation}
Comparing the residue at $u=\frac12$ with that for the pole
mass~\cite{BB,Bigi}, we obtain the ambiguity of the matching coefficients
due to the infrared renormalon at $u=\frac12$, in terms of the pole--mass
ambiguity, $\Delta m$:
\begin{equation}
\Delta C_\Gamma(\mu)=-\frac{1}{3}
\left[\frac{15}{4}+\eta(n-2)-2(n-2)^2\right]\frac{\Delta m}{m}\,,
\label{ir}
\end{equation}
at any scale, $\mu$.
This ambiguity must be compensated in~(\ref{1/m}) by
ultraviolet--renormalon ambiguities $\Delta G_1$
and $\Delta G_2$~\cite{NS}. Since~(\ref{ir}) is quadratic in
$\eta(n-2)$, we have 3 equations, with only 2 unknowns.
They are indeed consistent, and yield
\begin{equation}
\Delta G_1={\textstyle\frac34}\Delta m\,,\quad
\Delta G_2=-{\textstyle\frac16}\Delta m\,.
\label{G12}
\end{equation}
This result was obtained in~\cite{NS}, where the matching coefficients of
the $\eta=\pm1$ components of the $n=1$ current were considered. However,
the 2 ambiguities~(\ref{G12}) were obtained in~\cite{NS} from only 2
equations, with no consistency check. We have one check resulting
from one extra equation.

Finally, one might wonder what are the prospects of going to ${\rm
O}(1/\beta_0^2)$, where the renormalon structure is presumably far less
trivial, with the possibility of cuts appearing in the Borel transform,
rather than mere poles. The prospects appear to be rather slim, since one
would have to insert chains of light--quark loops into the gluon lines of
the diagrams of Fig.~\ref{Fig}. Even worse, one would need to insert a
loop into the three--gluon vertices. There is, however, one term that is
clearly tractable: that obtained by replacing the one--fermion--loop boson
self--energy of diagram~\ref{Fig}b by self--energy terms with $L$ loops, of
which $L-2$ are fermion loops. This was achieved in an ${\rm O}(1/N_f^2)$
analysis of the muon anomaly~\cite{LNF}, where the large--$L$ behaviour was
obtained, by evaluating an integral of the Borel transform of the
corresponding contributions to the Gell--Mann---Low function, also obtained
in closed form in~\cite{LNF}. Recently, corresponding contributions to the
HQET self--energy have been analyzed~\cite{Ben2}, with a result identical to
that in~\cite{LNF}.

\section{Summary}

Our principal result is formula~(\ref{cm}), which matches QCD and HQET
currents in~(\ref{RG}). It applies to all currents with a Dirac
structure $\Gamma=\gamma^{[\mu_1}\ldots\gamma^{\mu_n]}$
that anticommutes or commutes with $\gamma_0$.
The one--loop term agrees with~\cite{EH}. The coefficients
in~(\ref{coef}) and~(\ref{Delta}) give our new two--loop term,
for any gauge theory, current, and quark--mass ratios. To use the
general result one sets $n$ to the number of $\gamma$\/--matrices
in $\Gamma$ and uses $\eta=+(-1)^n$, or $\eta=-(-1)^n$,
according as whether $\Gamma$ anticommutes or commutes with $\gamma_0$.
Multiplying $\Gamma$ by a naively anticommuting $\gamma_5$ does not
change the matching coefficient.
Table~1 gives specific numerical results, ignoring small effects
of light/heavy quark--mass ratios.

All--order results, of similar generality, were obtained from the master
formula~(\ref{Fed}), which gives the ${\rm O}(1/\beta_0)$ terms in
the matching coefficients~(\ref{path}) and the anomalous
dimensions~(\ref{agj}). Infrared--renormalon ambiguities in the
matching coefficients are cancelled by the ultraviolet--renormalon
ambiguities~(\ref{G12}), derived in~\cite{NS} without benefit of our
consistency check.

We briefly note the following salient points.
\begin{enumerate}
\item Consistent results were obtained by including
heavy--quark--loop effects in all 6 of the terms of~(\ref{main});
one must not omit heavy loops from HQET.
\item The HQET on--shell renormalization constant~(\ref{Zos})
removes the non--uniformity, at zero light--quark mass, of the
corresponding QCD result~\cite{BGS}, as shown in~(\ref{RZ}).
\item The current--independent procedure of~(\ref{form}) and~(\ref{form1})
is an efficient alternative to evaluating traces with
complicated antisymmetrized products of $\gamma$\/--matrices.
\item Effects of non--trivial quark--mass ratios are obtained from the 3
integrals defined and evaluated in Appendix~A, where series expansions
are given.
\item Apart from such two--scale integrals, our calculations
were purely algebraic, as exemplified by the polynomial coefficients of
Appendix~B for the QCD vertex~(\ref{vert}).
\item All two--loop anomalous dimensions of QCD currents are given
by~(\ref{adim}), whose specialization~(\ref{adim2}) to the
tensor current appears to be new.
\item The matching coefficients~(\ref{cm})
confirm the universality of the
finite renormalizations~(\ref{Zp}) and~(\ref{Za})
that restore chiral symmetry to the 't~Hooft--Veltman
$\gamma_5$~\cite{Lar,LV}.
\item Two--loop corrections to ratios of meson decay constants are
given: in~(\ref{Rm}), which confirms the relation between
$\overline{\rm MS}$ and pole masses~\cite{GBGS};
in~(\ref{Rvp}), which gives a $-(3.1\pm0.5)$\% two--loop correction
to $f_{{\rm B}^*}/f_{\rm B}$, comparable to the $-(4.6\pm0.4)$\%
one--loop correction;
and in~(\ref{Rtv}), which gives the tensor coupling of B$^*$.
\item All our two--loop corrections are distressingly large; fastest
apparent convergence of $f_{{\rm B}^*}/f_{\rm B}$ would require
evaluating $\alpha_{\rm s}$ at $\mu=370$~MeV.
\item Infrared safety of ratios of matching coefficients, at finite
orders of perturbation theory, is exemplified by~(\ref{glum}),
which also exposes the infrared renormalon at $u=\frac12$.
\item Decoupling of the $t$\/--quark loop from
$f_{{\rm B}^*}/f_{\rm B}$ is explicitly demonstrated in~(\ref{deco}).
\item The $b$\/--quark loop approximately decouples
from $f_{{\rm B}^*}/f_{\rm B}$, as shown by~(\ref{approx}).
\item Naive non--abelianization of the massless--quark--loop contributions,
by the process $N_l\to N_l-\frac{33}{2}$, approximates our exact two--loop
terms, in ratios of decay constants, at the 30\% level, or better.
\item  In the $\overline{\rm MS}$ scheme, every large--$N_f$ series of the
type~(\ref{formal}) is formally resummed by~(\ref{path}), whose first
integral gives the scale--dependence, whilst the second contains
scheme--independent renormalon singularities.
\item Anomalous dimensions and renormalon residues,
for any current, are encoded by the $\varepsilon$\/-- and
$u$\/--dependencies of the master multinomial~(\ref{Ned}).
\item The anomalous current dimensions~(\ref{agj}) involve the
universal~\cite{BB,BK,PMP,LNF,JAG} $\varepsilon$--expansion~(\ref{epe}).
\item The anomalous field dimensions~(\ref{aqq}) confirm our
result $\tilde{\gamma}_J=\frac12\gamma_{\bar{q}q}+{\rm O}(1/\beta_0^2)$.
\item The ${\rm O}(1/m)$ contributions for any current matching are given
by~(\ref{1/m}).
\item The residue in~(\ref{Fu}) at $u=\frac12$ produces the
ambiguity~(\ref{ir}), which is cancelled in~(\ref{1/m}) by ambiguities in
the pole mass, $m$, and the matrix elements $G_{1,2}$.
\item The inclusion of ${\rm O}(1/\beta_0^2)$ terms in all--order
calculations will be very demanding, though some progress in the
abelian case has been made in~\cite{LNF,JAG,Ben2}.
\end{enumerate}

In conclusion: we hope that our results will serve both to enable
more accurate extraction of meson decay constants from lattice simulations
and sum rules, and also to provide detailed testing grounds for perturbative
and non--perturbative approximations based on large--$N_f$ expansions.

\noindent{\em Acknowledgments:} We thank the Royal Society and PPARC
for supporting our collaboration.
We are grateful for discussions with
P.~Ball, M.~Beneke, J.~A.~Gracey, N.~Gray, A.~C.~Hearn, A.~L.~Kataev,
S.~A.~Larin, C.~J.~Maxwell, M.~Neubert, C.~T.~Sachrajda, and K.~S.~Schilcher.
We thank D.~A.~Ross for organizing the 1994 UK High Energy Physics
Institute at Southampton, which was vital to the completion of this work.

\appendix

\newpage

\section{Dilogarithmic integrals $\Delta_k(r)$}
\label{Int} \setcounter{equation}{0}

The finite--mass corrections~(\ref{DeltaZ}) and~(\ref{DeltaG}) involve 3
integrals of the polarization operator subtracted at $m_i=0$. With
$z=-m_i^2/k^2$, this takes the form~\cite{GBGS}
\begin{equation}
\Pi(z) = 2(1-2z)\sqrt{1+4z}\arccoth\sqrt{1+4z} + \log z + 4z\,,
\label{Pim}
\end{equation}
which is then integrated over simple functions of
$y=2/(1+\sqrt{1-4m^2/k^2})$, to obtain
\begin{eqnarray}
&&\hspace{-11mm}
\Delta_1(r) = \frac{1}{6} \int\limits_0^1 {\rm d}y \frac{4-y}{1-y}
\Pi\left(r^2\frac{1-y}{y^2}\right)
= - (1+r)L_+(r) - (1-r)L_-(r) + \log^2 r + \zeta(2),
\nonumber\\
&&\hspace{-11mm}
\Delta_2(r) = \frac{2}{3} \int\limits_0^1 {\rm d}y
\Pi\left(r^2\frac{1-y}{y^2}\right)
= - r(1-r^2)L_+(r) + r(1-r^2)L_-(r) + 2r^2(\log r+1),
\label{Delta123}\\
&&\hspace{-11mm}
\Delta_3(r) = \frac{1}{6} \int\limits_0^1 {\rm d}y \, y
\Pi\left(r^2\frac{1-y}{y^2}\right)
= - r^3(1+r)L_+(r) + r^3(1-r)L_-(r) - r^2\left(\log r+\frac32\right),
\nonumber
\end{eqnarray}
in terms of the dilogarithmic integrals~\cite{GBGS,BGS}
\begin{equation}
L_\pm(r) = \int\limits_0^1 {\rm d}x \frac{\log x-\log r}{x\pm r}
= \left\{
\begin{array}{ll}
\frac12\log^2r - \log r \log(1\pm r) + \frac{1\mp3}2 \zeta(2) - \Li_2(\mp r),
& r\le1\,, \\
\log r \log\frac{r}{r\pm1} + \Li_2(\mp1/r), & r\ge1\,,
\end{array}
\right.
\label{Lpm}
\end{equation}
where $\Li_p(x)=\sum_{n=1}^\infty x^n/n^p$. Expanding~(\ref{Delta123}), for
$r<1$, we obtain
\begin{eqnarray}
&&\Delta_k(r) = 2\log r \sum_{n=1}^\infty g_k(n)r^{2n}
+ \sum_{n=1}^\infty g'_k(n)r^{2n} + \delta_k\,,
\label{smallr}\\
&&g_1(n) = \frac1{2n-1}-\frac1{2n}\,, \quad
\delta_1 = 3\zeta(2)r\,,
\nonumber\\
&&g_2(n) = \frac1{2n-1}-\frac1{2n-3}\,, \quad
\delta_2 = 3\zeta(2)r(1-r^2)\,,
\nonumber\\
&&g_3(n) = \frac1{2n-3}-\frac1{2n-4} \quad (n\ne2)\,, \quad
g_3(2)=1\,, \quad g'_3(2)=-2\,,
\nonumber\\
&&\quad \delta_3 = -r^4\log^2r
+ 3\zeta(2)r^3\left(1-{\textstyle\frac13}r\right)\,,
\nonumber
\end{eqnarray}
where $g'_k(n)={\rm d}g_k(n)/{\rm d}n$. Similarly, for $r>1$, we find
\begin{eqnarray}
&&\Delta_k(r) = -2\log r \sum_{n=0}^\infty \bar{g}_k(n)\frac1{r^{2n}}
+ \sum_{n=0}^\infty \bar{g}'_k(n)\frac1{r^{2n}} + \bar{\delta}_k\,,
\label{larger}\\
&&\bar{g}_1(n) = \frac1{2n}-\frac1{2n+1} \quad (n\ne0)\,, \quad
\bar{g}_1(0)=-1\,, \quad \bar{g}'_1(0)=2\,, \quad
\bar{\delta}_1 = \log^2r+\zeta(2)\,,
\nonumber\\
&&\bar{g}_2(n) = \frac1{2n+3}-\frac1{2n+1}\,, \quad
\bar{\delta}_2 = 0\,,
\nonumber\\
&&\bar{g}_3(n) = \frac1{2n+4}-\frac1{2n+3}\,, \quad
\bar{\delta}_3 = 0.
\nonumber
\end{eqnarray}
Finally, external--flavour contributions, with $m_i=m$, are obtained from
\begin{equation}
\Delta_1(1)=2\zeta(2)\,, \quad \Delta_2(1)=2\,, \quad
\Delta_3(1)=\zeta(2)-{\textstyle\frac32}\,.
\label{r1}
\end{equation}

\newpage

\section{QCD proper vertex}
\label{Vert}

Here we present all non--zero coefficients in the formula~(\ref{vert}) for
the QCD proper vertex. The \LaTeX{} source of these formulae has been
generated by a REDUCE program, using the library package RLFI by
R.~Liska~\cite{H}.
\begin{eqnarray}
&&a_{000}=
\left(3 d^{3}-30 d^{2}+101 d-110\right) \left(d-1\right) \left(d-2
\right) \left(d-5\right) \left(d-6\right)
\nonumber\\
&&a_{001}=
-2 \left(d^{5}-17 d^{4}+109 d^{3}-325 d^{2}+428 d-172\right) \left(d-5
\right) \left(d-6\right)
\nonumber\\
&&a_{002}=
-2 \left(5 d^{4}-50 d^{3}+164 d^{2}-156 d-56\right) \left(d-5\right)
\left(d-6\right)
\nonumber\\
&&a_{003}=
8 \left(d-4\right)^{2} \left(d-5\right) \left(d-6\right)
\nonumber\\
&&a_{010}=
\left(d^{2}-8 d+10\right) \left(2 d-7\right) \left(d-2\right) \left(d-4
\right) \left(d-5\right) \left(d-6\right)
\nonumber\\
&&a_{011}=
-2 \left(d^{3}-13 d^{2}+42 d-32\right) \left(d-4\right)^{2} \left(d-5
\right) \left(d-6\right)
\nonumber\\
&&a_{012}=
-2 \left(3 d^{2}-10 d-4\right) \left(d-4\right)^{2} \left(d-5\right)
\left(d-6\right)
\nonumber\\
&&a_{013}=
8 \left(d-4\right)^{2} \left(d-5\right) \left(d-6\right)
\nonumber\\
&&a_{030}=
-2 \left(7 d^{3}-76 d^{2}+255 d-246\right) \left(d-2\right) \left(d-3
\right) \left(d-4\right)
\nonumber\\
&&a_{031}=
4 \left(3 d^{4}-50 d^{3}+287 d^{2}-660 d+492\right) \left(d-3\right)
\left(d-4\right)
\nonumber\\
&&a_{032}=
32 \left(d^{2}-6 d+6\right) \left(d-3\right) \left(d-4\right)^{2}
\nonumber\\
&&a_{100}=
-\left(d^{4}-9 d^{3}+26 d^{2}-24 d+2\right) \left(3 d-8\right) \left(d+2
\right)
\nonumber\\
&&a_{101}=
4 \left(2 d^{6}-31 d^{5}+183 d^{4}-508 d^{3}+642 d^{2}-260 d-32\right)
\nonumber\\
&&a_{102}=
4 \left(2 d^{5}-2 d^{4}-139 d^{3}+794 d^{2}-1600 d+1088\right)
\nonumber\\
&&a_{103}=
-16 \left(2 d-7\right) \left(d-2\right) \left(d-4\right)
\nonumber\\
&&a_{110}=
-\left(2 d^{4}-24 d^{3}+105 d^{2}-199 d+134\right) \left(3 d-8\right)
\left(d-2\right)
\nonumber\\
&&a_{111}=
2 \left(2 d^{5}-33 d^{4}+210 d^{3}-643 d^{2}+916 d-448\right) \left(d-4
\right)
\nonumber\\
&&a_{112}=
2 \left(8 d^{4}-71 d^{3}+211 d^{2}-178 d-88\right) \left(d-4\right)
\nonumber\\
&&a_{113}=
-8 \left(2 d-7\right) \left(d-3\right) \left(d-4\right)
\nonumber\\
&&a_{120}=
4 \left(3 d-8\right) \left(d-2\right) \left(d-3\right) \left(d-4\right)
\nonumber\\
&&a_{121}=
-16 \left(d-2\right) \left(d-3\right) \left(d-4\right)^{2}
\nonumber\\
&&a_{122}=
-16 \left(d-2\right) \left(d-3\right) \left(d-4\right)
\nonumber\\
&&a_{200}=
-\left(d^{3}-12 d^{2}+45 d-46\right) \left(3 d-8\right) \left(d-6\right)
\nonumber\\
&&a_{201}=
2 \left(d^{3}-18 d^{2}+101 d-184\right) \left(d-2\right) \left(d-6
\right)
\nonumber\\
&&a_{202}=
8 \left(d^{3}-12 d^{2}+47 d-56\right) \left(d-6\right)
\nonumber\\
&&a_{230}=
2 \left(d^{2}-5 d+2\right) \left(3 d-8\right) \left(d-4\right)
\nonumber\\
&&a_{231}=
-4 \left(d^{3}-11 d^{2}+30 d-16\right) \left(d-4\right)
\nonumber\\
&&a_{232}=
-16 \left(d-2\right) \left(d-4\right)^{2}
\nonumber
\end{eqnarray}

\newpage
\raggedright

\end{document}